\documentstyle[12pt,fleqn,epsfig,axodraw]{article}

\newlength{\dinwidth}
\newlength{\dinmargin}
\newcommand{\smallPomeron}{\mbox{\scriptsize $I\!\!P$}}
\newcommand{\tinyPomeron}{\mbox{\tiny $I\!\!P$}}

\newcommand{\lsim}{\raise.25ex \hbox{ $<$ \kern-1.1em \lower1ex \hbox {$\sim$ }}}
\newcommand{\gsim}{\raise.25ex \hbox{ $>$ \kern-1.1em \lower1ex \hbox {$\sim$ }}}

\newcommand{\vm}{vector meson}

\newcommand{\vmp}{vector meson production}

\newcommand{\vmwf}{vector meson wave function}
\newcommand{\vmwfs}{vector meson wave functions}

\newcommand{\cs}{cross section}

\newcommand{\eqn}{equation }
\newcommand{\eqns}{equations }
\newcommand{\fig}{figure\ }
\newcommand{\figs}{figures\ }

\newcommand{\qq}{\mbox{$q \overline{q}$}}

\newcommand{\qqpair}{\mbox{$q \overline{q} $-pair}}
\newcommand{\qqp}{\mbox{$q \overline{q} $-pair}}

\newcommand{\qql}{\mbox{$q \overline{q} $-loop}}

\newcommand{\etal}{{\it et al.}}
\newcommand{\ie}{{\it i.e.}}

\newcommand{\pert}{perturbative}
\newcommand{\nonpert}{nonperturbative}

\newcommand{\nonrel}{non-relativistic}

\newcommand{\wf}{wave function}
\newcommand{\wfs}{wave functions}
\newcommand{\lc}{light-cone}
\newcommand{\lcwf}{light-cone wave function}
\newcommand{\lcwfs}{light-cone wave functions}

\newcommand{\PhiLsq}{\mbox{$\Phi \! \left( \bfLsq \right)$}}

\renewcommand{\r}{\mbox{$\rho$}}

\newcommand{\rmes}{\mbox{$\rho$\,-meson}}

\newcommand{\f}{\mbox{$\phi$}}

\newcommand{\fmes}{\mbox{$\phi$\,-meson}}

\newcommand{\jpsi}{\mbox{$J\! / \! \Psi$}}

\newcommand{\jpsimes}{\mbox{$J \! / \! \Psi$\,-meson}}

\newcommand{\pom}{pomeron}

\newcommand{\cm}{centre-of-mass}

\newcommand{\dep}{dependence}

\newcommand{\ra}{\mbox{$\rightarrow$}}

\newcommand{\qsq}{\mbox{$Q^{2}$}}

\newcommand{\gev}{{\rm Ge}\kern-1.pt{\rm V}}
\newcommand{\gevsq}{\mbox{$\mathrm{{\rm Ge}\kern-1.pt{\rm V}}^2$}}

\newcommand{\tinygev}{{\tiny \rm Ge}\kern-1.pt{\tiny \rm V}}
\newcommand{\tinygevsq}{\mbox{$\mathrm{{\tiny \rm Ge}\kern-1.pt{\tiny \rm V}}^2$}}

\newcommand{\kev}{{\rm ke}\kern-1.pt{\rm V}}
\newcommand{\kevsq}{\mbox{$\mathrm{{\rm ke}\kern-1.pt{\rm V}}^2$}}

\newcommand{\mev}{{\rm Me}\kern-1.pt{\rm V}}
\newcommand{\mevsq}{\mbox{$\mathrm{{\rm Me}\kern-1.pt{\rm V}}^2$}}

\newcommand{\vv}{\half V}

\newcommand{\mq}{m_{q}}
\newcommand{\mqsq}{\mbox{$m_{q}^{2}$}}

\newcommand{\mvsq}{\mbox{$M_{V}^{2}$}}

\newcommand{\beq}{\begin{equation}}
\newcommand{\eeq}{\end{equation}}

\newcommand{\beqarr}{\begin{eqnarray}}
\newcommand{\eeqarr}{\end{eqnarray}}

\newcommand{\aEM}{\mbox{$\alpha_{em}$}}

\newcommand{\aS}{\mbox{$\alpha_{\mbox{\tiny S}}$}}
\newcommand{\aSsq}{\mbox{$\alpha_{\mbox{\tiny S}}^2$}}

\newcommand{\ellt}{\mbox{$\ell_t$}}
\newcommand{\bfell}{\mbox{\boldmath $\ell$}}
\newcommand{\bfellt}{\mbox{\boldmath $\ell$}_t}
\newcommand{\bfelltsq}{\mbox{\boldmath $\ell$}_t^2}

\newcommand{\tildeellt}{\mbox{$\widetilde{\ellt}$}}

\newcommand{\bfL}{\mbox{$\mathbf L$}}

\newcommand{\bfLsq}{\mbox{${\mathbf L}^2$}}

\newcommand{\modbfL}{\mbox{$\left|{\mathbf L}\right|$}}

\newcommand{\bfktsq}{\mbox{\boldmath $k$}_t^2}

\newcommand{\tildekt}{\mbox{$\widetilde{k_t}$}}

\newcommand{\Deltat}{\mbox{${\Delta}_t$}}

\newcommand{\half}{{\textstyle {1 \over 2}}}

\newcommand{\quarter}{{\textstyle {1 \over 4}}}

\begin{document}
\begin{flushright}
M/C-TH-01/01 \\
\today \\
\end{flushright}
\begin{center}
\vspace*{2cm}

{\Large \bf A Unified Model of  Exclusive $\rho^0$, $\phi$ and $\jpsi$
Electroproduction}

\vspace*{1cm}

A Donnachie, J Gravelis and G Shaw
\footnote{{\tt ad@theory.ph.man.ac.uk, janis@theory.ph.man.ac.uk,
graham.shaw@man.ac.uk}}

\vspace*{0.5cm}
Department of Physics and Astronomy, University of Manchester,\\
Manchester, M13 9PL, England.
\end{center}
\vspace*{2cm}

\begin{abstract}

A two-component model is developed for diffractive electroproduction of 
$\rho^0$, $\phi$ and $\jpsi$, based on non-perturbative and perturbative 
two-gluon exchange. This provides a common kinematical structure for 
non-perturbative and perturbative effects, and allows the role of the 
vector-meson vertex functions to be explored independently of the 
production dynamics. A good global description of the vector-meson
data is obtained.

\end{abstract}


\newpage

\section{Introduction}

High-energy exclusive photo- and electroproduction of vector mesons offers a 
variety of insights into the diffractive mechanism. The choice of different 
vector mesons and a range of photon virtualities allows one to move from the
primarily nonperturbative regime to the primarily perturbative within one
framework, and to explore kinematical regions where neither is predominant. 
Vector-meson production also has the benefit of a high rate, but it has the 
disadvantage of dependence on the choice of the vertex functions
which couple the  vector mesons to the $q \bar{q}$ pairs. 
In both respects it differs from deep virtual Compton scattering, which is 
theoretically better defined but has the disadvantage of a small cross 
section.

On the basis of the factorisation theorem \cite{CFS97}, exclusive vector meson 
production can be considered as three separate processes: the fluctuation
of the (virtual) photon into a $q\bar{q}$ pair; the interaction of the
$q\bar{q}$ pair with the proton; and the formation of the vector meson 
from the $q\bar{q}$ pair which naturally involves the vector-meson vertex 
function. It may be argued that the structure of the nonpertubative 
vector-meson vertex function invalidates the proof of the factorisation
theorem as it leads to additional contributions. However it has been
shown \cite{HL95}, at least in a simple model for the vertex function,
that gauge invariance ensures that the additional contributions cancel 
and factorisation is preserved. 

The aim of this paper is twofold: to obtain a global description of 
exclusive vector meson photo- and electroproduction; and to explore
the  choice of vertex function.

The interaction of the $q\bar{q}$ pair with the pomeron is modelled by 
two-gluon exchange, figure 1, which can be applied both to nonperturbative 
and perturbative gluon exchange. The former is based on the model of Diehl 
\cite{MD:1995} and the latter either by utilising the gluon structure function
or following the model of Cudell and Royon \cite{Cudell_Royen:1998}. 

The approach has the 
advantage of providing a common kinematical structure in which it is possible 
to separate the part of the vector-meson production amplitude describing the 
kinematics from the part which describes the dynamics of the process 
\cite{DGS00}. This separation then allows the nonpertubative and perturbative 
contributions to be combined at the amplitude level. Thus the approach 
follows recent ideas about two pomerons and two-component models of 
diffraction \cite{KMS:95_97} through \cite{GLMN99}, combining 
``soft'' (nonperturbative) and ``hard'' (perturbative) terms. This is 
essential for a global approach as neither a nonperturbative nor a 
perturbative model alone can describe  all the observed 
features of diffractive vector-meson  photo- and electroproduction. 

In a previous paper \cite{DGS00} we have successfully applied this 
approach  to calculate
$\rho$ electroproduction in a model which avoids the vertex function
complication, following \cite{Martin_Ryskin_Teubner:1996}, by considering 
open $q\bar{q}$-pair production 
with the invariant mass of the $q\bar{q}$ pair restricted to the region of 
the $\rho$ mass. However this method cannot be applied to higher-mass 
vector mesons:  for a $q\bar{q}$ with an invariant mass in the region
of the $\rho$ there are no states available except the $\rho$ itself
(making a $10\%$ allowance for $\omega$ production), but this is not 
true for the $\phi$ and $J/\psi$. To obtain a global description of $\rho$,
$\phi$ and $J/\psi$ photo- and electroproduction a kinematical framework
involving vector-meson vertex functions has to be used. 

In figure 1 the quark lines marked with a cross are off-shell, and the
vertex function must necessarily take this into account. We use the 
prescription of Cudell and Royon \cite{Cudell_Royen:1998} for these vertex functions.
They appear very different from the usual on-shell vertex functions, for
example those of Brodsky and Lepage \cite{brodsky_etal_light_cone_pert_theory:1980}.
However  they are both 
obtained by a boost of the same vertex function from the vector-meson 
centre-of-mass and differ only by the imposition or relaxation of the 
on-shell condition.

Kinematical and dynamical considerations of the two-gluon exchange model
are coverd in sections 2 and 3 and the vector-meson vertex functions are
discussed in section 4. The model is applied to the data in section 5, 
and conclusions are given in section 6.

\begin{figure}[!!!t]
\begin{center} 
%
\begin{picture}(340,70)(20,95)
\Photon(30,152)(90,152){5}{5}
%
\Vertex(90,152){2}
\Curve{(90,152)(95,143)(98,142)}
\Curve{(90,152)(95,161)(98,162)}
%
\ArrowLine(283,162)(190,162)
\Line(190,162)(97,162)
\Line(97,142)(190,142)
\ArrowLine(190,142)(283,142)
%
\Line(129,138)(137,146)
\Line(129,146)(137,138)
%
\Line(246,138)(254,146)
\Line(246,146)(254,138)
%
\Line(91,78)(289,78)
\Line(91,80)(289,80)
\Line(91,82)(289,82)
\Line(69,80)(89,80)
\Line(291,80)(311,80)
%
\GOval(90,80)(4,2)(0){1}
\GOval(290,80)(4,2)(0){1}
%
\Vertex(170,83){2}
\Vertex(170,142){2}
\Vertex(210,83){2}
\Vertex(210,142){2}
%
\Gluon(170,83)(170,142){5}{6}
\Gluon(210,83)(210,142){5}{6}
%
\LongArrow(162,100)(162,124)
\Text(157,112)[cr]{$k$}
%
\LongArrow(218,124)(218,100)
\Text(223,112)[cl]{$k - \Delta$}
%
\DashLine(180,180)(190,180){4}
\DashLine(190,180)(190,70){4}
\DashLine(190,70)(200,70){4}
\Text(170,150)[cc]{$\mu$}
\Text(210,150)[cc]{$\nu$}
%
\ArrowLine(290,151)(350,151)
\ArrowLine(290,153)(350,153)
%
\Curve{(290,152)(285,143)(282,142)}
\Curve{(290,152)(285,161)(282,162)}
\GOval(286,152)(10,4)(0){0.5}
%
\LongArrow(306,170)(293,157)
\Text(308,172)[lb]{$\Phi_{q \overline{q} \ra V}$}
\Text(20,152)[r]{$q, \ \epsilon$}
\Text(59,80)[r]{$p$}
\Text(321,80)[l]{$\widetilde{p}$}
\Text(290,130)[cr]{$\frac{1}{2} V + \ell$}
\Text(290,174)[cr]{$- \left[ \frac{1}{2} V - \ell \right]$}
%
%
\Text(170,75)[ct]{$\alpha$}
\Text(210,75)[ct]{$\beta$}
\Text(360,150)[l]{$V, \ e$}
\Text(8,112)[cc]{${\mathbf (a)}$}
\end{picture}
%
%
%
%
\begin{picture}(340,160)(20,80)
\Photon(30,152)(90,152){5}{5}
%
\Vertex(90,152){2}
\Curve{(90,152)(95,143)(98,142)}
\Curve{(90,152)(95,161)(98,162)}
%
\ArrowLine(283,162)(190,162)
\Line(190,162)(97,162)
\Line(97,142)(190,142)
\ArrowLine(190,142)(283,142)
%
\Line(129,158)(137,166)
\Line(129,166)(137,158)
%
\Line(246,138)(254,146)
\Line(246,146)(254,138)
%
\Line(91,78)(289,78)
\Line(91,80)(289,80)
\Line(91,82)(289,82)
\Line(69,80)(89,80)
\Line(291,80)(311,80)
%
\GOval(90,80)(4,2)(0){1}
\GOval(290,80)(4,2)(0){1}
%
\Vertex(170,83){2}
\Vertex(170,162){2}
\Vertex(210,83){2}
\Vertex(210,142){2}
%
\Gluon(170,83)(170,162){5}{8}
\Gluon(210,83)(210,142){5}{6}
%
\LongArrow(162,100)(162,124)
\Text(157,112)[cr]{$k$}
%
\LongArrow(218,124)(218,100)
\Text(223,112)[cl]{$k - \Delta$}
%
\DashLine(180,180)(190,180){4}
\DashLine(190,180)(190,70){4}
\DashLine(190,70)(200,70){4}
\Text(170,170)[cc]{$\mu$}
\Text(210,150)[cc]{$\nu$}
%
\ArrowLine(290,151)(350,151)
\ArrowLine(290,153)(350,153)
%
\Curve{(290,152)(285,143)(282,142)}
\Curve{(290,152)(285,161)(282,162)}
\GOval(286,152)(10,4)(0){0.5}
%
\LongArrow(306,170)(293,157)
\Text(308,172)[lb]{$\Phi_{q \overline{q} \ra V}$}
\Text(170,75)[ct]{$\alpha$}
\Text(210,75)[ct]{$\beta$}
\Text(20,152)[r]{$q, \ \epsilon$}
\Text(59,80)[r]{$p$}
\Text(321,80)[l]{$\widetilde{p}$}
\Text(290,130)[cr]{$\frac{1}{2} V + \ell$}
\Text(290,174)[cr]{$- \left[ \frac{1}{2} V - \ell \right]$}
\Text(360,150)[l]{$V, \ e$}
\Text(8,112)[cc]{${\mathbf (b)}$}
%
%
\end{picture}
\end{center}
\caption{
Two of the four \vmp\ diagrams in the kinematical framework of Cudell
and Royen \cite{Cudell_Royen:1998}.  The other two diagrams differ by
reversal of the quark charge flow and give the same contribution to
the \cs.
The off-shell quarks are marked by crosses, dashed lines indicate cuts
along which the quark lines are put on-shell, the minus sign indicates
the momentum of an antiparticle.
\label{CR_advanced_model:main_diagram}
}
\end{figure}
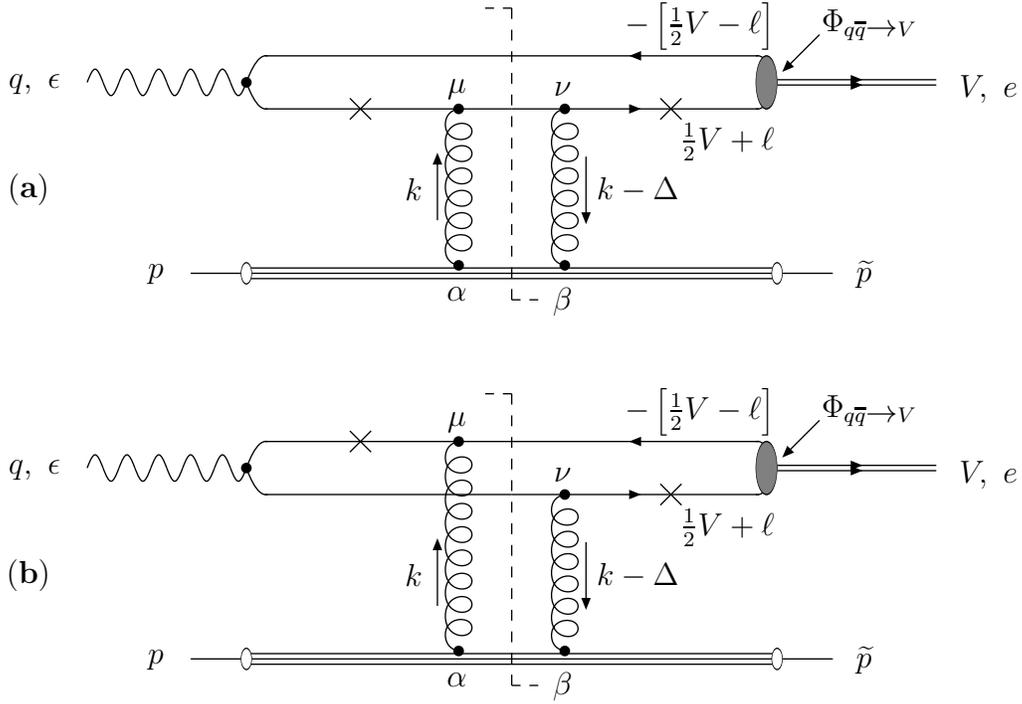

\section{The Kinematical Framework}

 The kinematical framework of our gluon exchange models
is depicted in \fig\ref{CR_advanced_model:main_diagram} and was
developed by Cudell and Royen \cite{Cudell_Royen:1998}.
As noted earlier, it
 involves the vector-meson vertex function $\Phi \left( \ell \right)$ 
with  either the quark or the anti-quark  off-shell.
The various four momenta used in our discussion are also defined 
in \fig\ref{CR_advanced_model:main_diagram}.  In terms of them,   the photon
virtuality $Q^2 = -q^2$
 and  the squared centre of mass energy of
the  photon-proton pair $W^2 = (p + q)^2$. 
The traces corresponding to the \qql\ in  diagrams (a) and (b) of
\fig\ref{CR_advanced_model:main_diagram} are  given by
\beqarr
{\mathcal T}_{\mathbf a}^{\mu \nu} &=& Tr 
\Big\{ 
\Phi \left( \ell \right) \left( \gamma . e \right)
\left( \mq + \gamma . \left[ \ell + \vv \right] \right)
{\gamma}^{\nu}
\left( \mq + \gamma . \left[ q + k + \ell - \vv \right] \right)
\nonumber
\\
&\times& 
{\gamma}^{\mu}
\left( \mq + \gamma . \left[ q + \ell - \vv \right] \right)
\left( \gamma . \epsilon \right)
\left( \mq + \gamma . \left[ \ell - \vv \right] \right)
\Big\}
\label{eqn:Tr_CR_a}
\\
%
%
{\mathcal T}_{\mathbf b}^{\mu \nu} &=& Tr 
\Big\{ 
\Phi \left( \ell \right) \left( \gamma . e \right)
\left( \mq + \gamma . \left[ \ell + \vv \right] \right)
{\gamma}^{\nu}
\left( \mq + \gamma . \left[ q + k + \ell - \vv \right] \right)
\nonumber
\\
&\times& 
\left( \gamma . \epsilon \right)
\left( \mq + \gamma . \left[ k + \ell - \vv \right] \right)
{\gamma}^{\mu}
\left( \mq + \gamma . \left[ \ell - \vv \right] \right)\Big\} ~.
\label{eqn:Tr_CR_b}
\eeqarr
One of the two quark lines emerging from the ${\gamma}^{*} \qq$ vertex
is off-shell with different virtualities in each of the diagrams of
\fig\ref{CR_advanced_model:main_diagram}, giving 
\beqarr
P_{\mathbf a} 
&=& \left[ q + \ell - \vv \right]^{2} - \mqsq 
\ = \ \ell . \ell + 2 \ell . q - \ell . V - q.V - Q^2 - \mqsq + \quarter \mvsq
\label{eqn:CR_adv_prop_a}
\\
\nonumber
\\
P_{\mathbf b} 
&=& \left[ k + \ell - \vv \right]^{2} - \mqsq
\ = \ k.k + 2 k . \ell - k.V + \ell . \ell - \ell . V - \mqsq + \quarter \mvsq 
\label{eqn:CR_adv_prop_b}
\eeqarr
for the denominators of the corresponding propagators.
Here we assume the diffractive amplitudes are completely dominated by their 
imaginary parts, which are  evaluated using the cuts shown 
by the dashed lines in  \fig\ref{CR_advanced_model:main_diagram}. 
The sum of both diagrams for the \qql\ gives
\beq
{\mathcal T}^{\mu \nu} =
\frac{1}{P_0} 
\left[ 
\frac{{\mathcal T}_{\mathbf a}^{\mu \nu}}{P_{\mathbf a}} 
+ \frac{{\mathcal T}_{\mathbf b}^{\mu \nu}}{P_{\mathbf b}}
\right] \,
\left( 2 \pi \right)^2
\delta \left( \left[ q + k + \ell - \vv \right]^2 - \mqsq \right) \
\delta \left( \left[ \ell - \vv \right]^2 - \mqsq \right)
\label{eqn:CR_adv_T_alpha_beta}
\eeq
where the $\delta$-functions are due to the on-shell conditions along
the cuts of the quark and the anti-quark lines respectively.  The
denominator
\beqarr
P_0 \ = \
\left[ \ell + \vv \right]^2 - \mqsq
\ = \ 2 {\ell}^2 - 2 m_q^2 + \half \mvsq
\label{eqn:CR_adv_P0}
\eeqarr
of the propagator for the off-shell quark forming
the vector meson 
is the same for both diagrams of \fig\ref{CR_advanced_model:main_diagram}.
In the Regge limit the proton line gives a contribution  $4
p^{\alpha} p^{\beta}$ in amplitude and the intermediate proton state is
cut and its mass neglected, yielding $\delta \left( \left[ p - k
\right]^2 \right)$.
The rest of the diagram, including the gluon propagators and the
description of the interaction of the pomeron with the proton, is
contained within the dynamical part $\mathcal P$, which  is
model-dependent.
Formally it is the (gauge-dependent) gluon propagator that contracts
the indices at $p_{\alpha} p_{\beta}$ and ${\mathcal T}_{\mu \nu}$.
Practically, the leading contribution in the Regge region  comes
from $g^{\alpha \mu} \, g^{\beta \nu}$ in the gluon propagators. Details of 
this and of
the decompositions of the four-vectors in terms of $p$ and $q$ for the
Regge region, where $W$ is significantly greater than any other scale
present, can be found in  \cite{{Cudell_Royen:1998},{JG_thesis:2000}}.
Here we would only
like to note the following points.

It is convenient to use the \lc\ variables 
 $P_t = \left( P^+, \,
{\mathbf P_t}, \, P^- \right)$ in
decompositions and in the further derivation, where
 $P^{\pm} \equiv P^0 \pm P^3$ and
the two-vector ${\mathbf P_t}$ lies in the
transverse plane, defined as the plane perpendicular
to the $\gamma^{*} p$ axis.
The variable $z$  is also often used. It is 
the fraction of the ``$+$'' momentum
of the photon carried by the quark, so that  $P_q^{+} \equiv z q^{+}$,
implying  $P_{\overline{q}}^{+} = \left( 1 - z
\right) q^{+}$ for the anti-quark.
The decomposition of the gluon four-momenta $k$ and $\left[ k - \Delta
\right]$ show \cite{Cudell_Royen:1998} that the gluon 
four-momenta are predominantly transverse,
$|k^2| \approx \bfktsq$.

For clarity  we rewrite \eqn (21) of
\cite{Cudell_Royen:1998} using our notation with the dynamical part 
$\mathcal P$ and the \lc\ variable $z$:
\beqarr
{\mathcal A}^{L, \, Tr} &=&
\frac{2}{3} \, \left( 4 \pi \right)^2 {\mathrm f}_q \sqrt{4 \pi \aEM}
\int \!\! \frac{d^2 \tildekt}{(2\pi)^2} 
%
%
%
\int \!  \frac{2 \, d z \, d^2 \tildeellt }{(2\pi)^3} 
\ \frac{1}{\sqrt{3}} \ \Phi \! \left(z,  | \tildeellt |  \right)
\ 3 \, {\mathcal P} \! \left( k, \Delta \right) \
\frac{N^{L, \, Tr}}{D}
%
%
\label{eqn:CR_offshell_final_amplitude}
\eeqarr
where we have introduced 
 $\tildeellt, \; \tildekt $  for  twice the transverse
parts  $\ellt, \,  k_t$ of the four-vectors $\ell, k$.  For \f\ and \jpsi\ 
electroproduction,  $\: {\mathrm f}_q = -\frac{1}{3}, \; \frac{2}{3}$ 
corresponding to the charge of the quark forming the
\vm\, while  the linear combination of the
$u\overline{u}$ and $d\overline{d}$ quark anti-quark pairs forming the
\rmes\ gives ${\mathrm f}_q = \frac{1}{\sqrt{2}}$.
The expressions for $N^L$ and $N^{Tr}$ are given in
\cite{Cudell_Royen:1998} and
%
%
\beqarr
%
D &=& 16
\left( 2 (2z-1) \ \Deltat . \tildeellt - \tildeellt . \tildeellt + 
4 \mqsq - 4z\left( 1 - z \right) \mvsq - t (2z-1)^2 \right)
\nonumber
\\
&\times&
\left( 4z \left( 1 - z \right) \qsq + 4 \mqsq - 
t - \left[ \tildeellt + \tildekt - 2 \Deltat \right] . 
\left[ \tildeellt + \tildekt \right] \right)
\nonumber
\\
&\times&
\left( 4z \left( 1 - z \right) \qsq + 
\left[ 2 \Deltat - \tildeellt \right] . \tildeellt + 4 \mqsq - t \right)~.
\label{eqn:CR_offshell_final_denominator}
\eeqarr 
%
%
The first line in $D$ originates from
%
%
the denominator of the off-shell quark propagator
$P_0$, \eqn(\ref{eqn:CR_adv_P0}); similarly,
%
%
the second and third lines are proportional to  $P_{\mathbf b}$ and 
$P_{\mathbf a}$ respectively, \eqns (\ref{eqn:CR_adv_prop_b}) and
(\ref{eqn:CR_adv_prop_a}).
The differences from the expression for $D$ given in
\cite{Cudell_Royen:1998} are due to the fact that we attribute the
gluon propagators to the dynamical part of the pomeron exchange.

Assuming $s$-channel helicity conservation the differential \cs\
is given by
\beq
\frac{d \sigma}{dt} \ = \ 
\frac{d \sigma^{Tr}}{dt} + \varepsilon_{expt} \, \frac{d \sigma^{L}}{dt} 
\ = \ \frac{1}{16 \pi W^4} 
\left( \left| {\mathcal A}^{Tr} \right|^2 +
\varepsilon_{expt} \left| {\mathcal A}^{L} \right|^2 
\right)
\eeq
where the polarisation of the photon beam $\varepsilon_{expt}$ is a
known characteristic of the experiment.  For HERA, $\varepsilon_{expt}
\approx 1$. For fixed-target experiments, it typically lies in the
range 0.5 to 0.9 depending on the energy and photon virtuality. 

Finally we note that the kinematical expressions  in this derivation
and in \cite{Cudell_Royen:1998} are valid for the pair of
diagrams depicted in \fig\ref{CR_advanced_model:main_diagram}.  The
other two diagrams differ by reversal of the quark charge flow and
thus have different traces, cut conditions and propagators than 
(\ref{eqn:Tr_CR_a})--(\ref{eqn:CR_adv_P0}) leading to different
decompositions of the four-vectors $\ell$ and $k$.
However, the only net difference is a change of sign in front of  $\ell$
%
%
in all expressions.  Because the integral over $\ellt $ is two dimensional,
(\ref{eqn:CR_offshell_final_amplitude}) gives the same answer
regardless of the sign in front of $\ell$ and the additional diagrams
result in a factor of 2 in front of the final expression for the amplitude.
%


\section{The Dynamics of Pomeron Exchange}

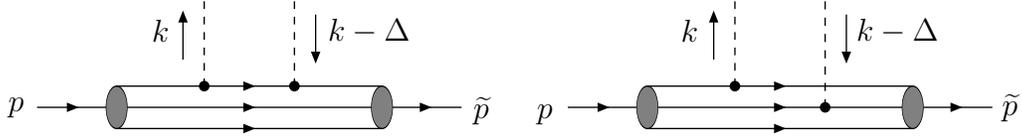
\begin{figure}[!!!t]
\begin{center} 
%
%
\begin{picture}(350,60)(0,10)
%
%
%
\ArrowLine(0,20)(27,20)
\ArrowLine(30,12)(130,12)
\ArrowLine(30,20)(130,20)
\ArrowLine(30,28)(130,28)
\ArrowLine(133,20)(160,20)
\GOval(30,20)(8,4)(0){0.5}
\GOval(130,20)(8,4)(0){0.5}
\DashLine(63,28)(63,60){3}
\Vertex(63,28){2}
\DashLine(97,28)(97,60){3}
\Vertex(97,28){2}
%
\LongArrow(55,38)(55,55)
\Text(50,49)[cr]{$k$}
%
\LongArrow(105,55)(105,38)
\Text(110,49)[cl]{$k - \Delta$}
%
%
%
%
\ArrowLine(200,20)(227,20)
\ArrowLine(230,12)(330,12)
\ArrowLine(230,28)(330,28)
\ArrowLine(230,20)(330,20)
\ArrowLine(333,20)(360,20)
\GOval(230,20)(8,4)(0){0.5}
\GOval(330,20)(8,4)(0){0.5}
\DashLine(263,28)(263,60){3}
\Vertex(263,28){2}
\DashLine(297,20)(297,60){3}
\Vertex(297,20){2}
\Text(-5,20)[r]{$p$}
\Text(165,16)[lb]{$\widetilde{p}$}
\Text(195,16)[rb]{$p$}
\Text(365,20)[l]{$\widetilde{p}$}
%
\LongArrow(255,38)(255,55)
\Text(250,49)[cr]{$k$}
%
\LongArrow(305,55)(305,38)
\Text(310,49)[cl]{$k - \Delta$}
\end{picture}
\end{center}
\caption{The two possibile ways in which the two gluons can couple to
the valence quarks of the proton.
In the \nonpert\ LN approach as applied by Diehl \cite{MD:1995} it is
argued \cite{LN:1987} that the diagrams in which the gluons couple to
different quarks are suppressed and subsequently can be neglected:
see text.
\label{gluon_gluon_proton_coupling}
}
\end{figure}

We investigate two ``summation models:'' these include
both ``hard''(perturbative) and ``soft''(nonperturbative)
components, since neither alone can  account for all the data. 
Before describing these summation models, we first summarise the models for 
the hard and soft terms  on which they are based. 

\subsection{The ``soft term''}

For this contribution we use
the \nonpert\ approach of Diehl \cite{MD:1995} based on earlier work by
 Landshoff and
Nachtmann (LN) \cite{LN:1987}.
In it the gluons are assumed not to interact with each other and a \nonpert\
gluon propagator \cite{MD:1995} 
\beq
{\mathcal D}_{np} \left( - k^2 \right) \ = \ 
{\mathcal N}_{np} 
\left[ 1 + \frac{k^2}{\left( n - 1 \right) \mu_0^2 } \right]^{-n}
\label{eqn:nonpert_propagator}
\eeq
is used with $n = 4$. The normalisation ${\mathcal N}_{np}$ is determined from
the condition
\beq
\int_{0}^{\infty} dk^{2} \left[ \alpha_{S}^{(0)} 
{\mathcal D}_{np}(k^{2}) \right]^{2} 
\ = \ \frac{9 \beta_{0}^{2}}{4 \pi} ~.
\label{eqn:ln_prop_normalisation_1}
\eeq
The phenomenological parameters $\beta_{0}$, which describes the
effective coupling of the pomeron to the proton, and $\mu_{0}$ are
determined from the total $pp$ and $p \overline{p}$ \cs\ data and from
deep inelastic scattering: $\beta_{0} \approx 2.0 \ \gev^{-1}$ and
$\mu_{0} \approx 1.1 \ \gev$ \cite{DL:1988_1989}. For the
nonperturbative couplings of the gluons to the quarks forming the
$\rho$ a value $\alpha_{S}^{(0)} \approx 1$ is taken. Of course the
precise value of $\alpha_{S}^{(0)}$ cannot be strictly specified,
and as we shall discuss in section 5 there is some flexibility
through the interplay with the choice of vertex function for the 
vector meson.

Landshoff and Nachtmann  \cite{LN:1987} have argued  that  diagrams in which 
the nonperturbative gluons couple to different valence quarks in the
proton, as shown in \fig \ref{gluon_gluon_proton_coupling}, are suppressed and
can be disregarded.
Hence only the diagrams where both gluons couple to the same valence quark
are calculated.  Each of the three valence quarks is incorporated into
the proton using the Dirac form factor
$F_{1p}(t)$, where $t = \Delta^2$.
The energy \dep\ of the soft \pom\ comes via a factor
$x_{\tinyPomeron}^{-\alpha_{\tinyPomeron}(t)}$ in the amplitude, where
\beq
x_{\tinyPomeron} \ \equiv \
\frac{M_V^{2} + Q^{2} - t }{ W^{2} + Q^{2} - m_{proton}^{2}}
\label{eqn:x_pom_definition}
\eeq
and $\alpha_{\tinyPomeron}(t) = 1.08 + 0.25 \, t$ is the soft \pom\
trajectory \cite{DL:1986_pomeron_intercept}.
The coupling $\aS$ at both vertices at the proton end is taken at a
\nonpert\ scale, \ie\ $\alpha_S^{(0)}$ is used.
For the vertices where the gluons couple to an off-shell quark line the
coupling is taken at a \pert\ scale $\lambda^2 = \left(
\bfelltsq + \mqsq \right) \left( \qsq + M_V^2 \right) / M_V^2$ which,
as argued in \cite{MD:1995}, is a typical scale for the whole upper
part of the diagram.
Thus for the dynamical part of the \nonpert\ approach one has
\beqarr
{\mathcal P}_{np} &=&
F_{1p}(t) \ 
x_{\tinyPomeron}^{ 1 - \alpha_{\tinyPomeron} \left( t \right) } \
{\alpha}_S^{(0)} \ \aS \! \left( \lambda^2 \right) \
{\mathcal D}_{np} \left( - k^2 \right) \
{\mathcal D}_{np} \left( - \left[ k - \Delta \right]^2 \right)~.
\label{eqn:offshell_nonpert_prop_related_expr}
\eeqarr
This term alone gives an energy \dep\ that is too flat
at the higher values of $Q^2$ due to the soft \pom\ intercept. 


\subsection{The ``hard term''}

Two alternative models, the standard perturbative QCD approach and 
the Cudell-Royon approach, will be considered.

\subsubsection{The ``standard'' perturbative model}

The perturbative QCD approach  can only be calculated
at $t=0$, and is based on the ideas of Martin \etal\
\cite{Martin_Ryskin_Teubner:1996} who applied it to  
\rmes\  electroproduction. In it
the pomeron is modelled as a pair of perturbative gluons, using the
perturbative gluon propagator ${\mathcal D}_p \! \left( k^2 \right) =
1 / k^2$.
The gluons are considered as part of the proton so that there are no
$\sqrt{\aS}$ couplings for the two bottom vertices.
In principle the gluon flux can be obtained from the unintegrated
gluon density $f ( x_{\tinyPomeron}, \bfktsq )$, which gives the
probability of finding a $t$-channel gluon with the momentum squared
$\bfktsq$ in the proton. However, a special treatment of the infrared
region is required because the unintegrated gluon density $f
(x_{\tinyPomeron}, \bfktsq)$ is  undefined as $\bfktsq
\to 0$ and numerically unavailable below some value of $\bfktsq =
Q_0^2$, which varies with the parton distribution chosen and 
is typically in the region from 0.2 to a few \gevsq.  The linear approximation
as suggested in \cite{Martin_Ryskin_Teubner:1996} is used to account
for the contribution to the integral from the $\bfktsq < Q_0^2$
region. This procedure has no direct physical significance. It serves
only to provide a continuous integrand and acts as a means of normalisation
of the perturbative contribution. A simple cutoff at an appropriate $Q_0^2$
would be equally effective but somewhat less elegant. 

As no direct physical significance can be attached to the contribution 
from this infrared part of the perturbative term there
 is not an element of double counting.
The separation between ``perturbative'' and ``nonperturbative'' is given
uniquely by the energy dependence of the two contributions. An implication of
this approach is that the perturbative (hard) term can contribute at $Q^2 = 0$, 
which is a feature of two-component models.

Thus for the dynamical part of the \pert\ approach one has
\beqarr
{\mathcal P}_{p} &=&
\frac{\pi}{4} \
\sqrt{\aS \! \Big( k^2 \Big) \ \aS \! \left( \left[ k - \Delta \right]^2 \right)} 
\ \frac{f \! \left( x_{\smallPomeron}, 
\sqrt{k^2 \ \left[ k-\Delta \right]^2}
\right) 
}{k^2 \ \left[ k-\Delta \right]^2}
\label{eqn:offshell_pert_prop_related_expr}
\eeqarr
where $f( x_{\tinyPomeron}, \bfktsq)$ is related to the gluon distribution
$g(x_{\tinyPomeron},Q^2)$ by 
\beq
x_{\tinyPomeron} \, g(x_{\tinyPomeron},Q^2) \ = \ 
\int^{Q^2}{{d\bfktsq}\over{\bfktsq}}
\, f( x_{\tinyPomeron}, \bfktsq)
\label{ftog}
\eeq
with the inverse
\beq
f( x_{\tinyPomeron}, \bfktsq) \ = \ 
\bfktsq \, {{\partial\Big(x_{\tinyPomeron}, 
g(x_{\tinyPomeron},\bfktsq)\Big)}\over{\partial\bfktsq}}~.
\label{gtof}
\eeq

This applies at
at $t=0$, and the experimental slope or some other ansatz must be used 
to compare with the integrated \cs.
Here we merely note  that this term alone gives an  energy
\dep\ which is clearly too steep for much of the data.
%

\subsubsection{The Cudell--Royen (CR) model}

Cudell and Royen \cite{Cudell_Royen:1998} propose a somewhat
different approach to the ``hard'' contribution.
Again perturbative gluon propagators of the form  $1/k^2$ are used.
However, since the \cs\ is not infrared divergent, it
is suggested that the divergencies from  diagrams where both gluons
couple to the same quark are cancelled in the infrared limit by
the divergencies from the diagrams where each gluon couples to a
different quark, \fig\ref{gluon_gluon_proton_coupling}.  In terms
of  form factors, the former case is described by the Dirac form
factor of the proton $F_{1}(t)$; while for the latter the form factor
\beq
{\mathcal E}_{2} \left( k, k-\Delta \right) 
\ = \ F_{1} \left( k^{2} + \left[ k-\Delta \right]^{2} 
+ c \ k . \left[ k-\Delta \right] \right) \; ,
\label{eqn:CR_epsilon_2}
\eeq
depending on the momenta of both gluons, is applied
with $c \approx -1$ as suggested by Cudell and Nguyen
\cite{Cudell_Nguyen_formfactors:1994}.
Unlike the ``standard'' perturbative approach, this approach does not contain 
any energy \dep\, but describes the $t$-dependence at a fixed energy. The
energy dependence has
to be introduced by hand, just as it was for the nonperturbative term, via
a factor $x_{\tinyPomeron}^{1-\alpha_{\tinyPomeron_0}}$. We assume a flat 
hard-pomeron trajectory $\alpha_{\tinyPomeron_0}=1.44$, independent of $t$.
Further the overall normalization is not uniquely specified as it is not
obvious what value of $\alpha_s$ should be used for coupling perturbative
gluons to bound quarks. 
%
%
Cudell and Royon \cite{{Cudell_Royen:1998},{Cudell_Royen:1996}} introduced an
effective factor ${\mathcal R}$ in the \cs\  with
 ${\mathcal R} \aSsq = 0.6$
\cite{{Cudell_Royen:1998},{Cudell_Royen:1996}}, a procedure which we adopt here.
In this way we finally obtain
\beq
{\mathcal P}_{CR} \ = \
\sqrt{ \Big\{ {\mathcal R} \aSsq \Big\} \ \aS \Big( k^2 \Big) \
\aS \left( \left[ k - \Delta \right]^2 \right)}
\ \frac{
F_{1}(t) - {\mathcal E}_{2} \left( k, k-\Delta \right)}
{k^2 \ \left[ k-\Delta \right]^2}x_{\tinyPomeron}^{1-\alpha_{\tinyPomeron_0}}~.
\label{eqn:P_CR_general_form}
\eeq


\subsection{The Summation Models}

Here we suggest two ``summation models'' which combine both hard and soft 
terms at the amplitude level in order to obtain a global description 
of \vmp\ data.

\subsubsection{Summation model S1}

This model is based on our earlier work \cite{DGS00}, which  modelled
\r\ electroproduction by ``open pair'' production in the region of the \r\ 
mass.
In particular, a successful description of the  data was obtained 
by combining the nonperturbative amplitude of section 3.1 with the
perturbative amplitude of section 3.2.1, using an empirical slope
parameter to describe the $t$-dependence of the latter.
In doing so,  we exploited several gluon distributions to calculate
the perturbative contribution using the PDFLIB program libraries
\cite{PDFLIB_manual} for numerical calculations.  However, 
it was found  that the best fit to the \rmes\ electroproduction
data, and especially to the energy \dep\ of the production \cs, was
obtained using the CTEQ4LQ \cite{CTEQ} gluon distribution.  We
continue to use the CTEQ4LQ gluon distribution in the present paper.

In this paper we explicitly incorporate vertex function effects using 
(\ref{eqn:CR_offshell_final_amplitude}) in order to treat  
the  \f\ and \jpsi\ as well as the \r\ electroproduction
in a common framework. To do this, we again need to extend the 
perturbative amplitude to $t \ne 0$. 
Since the proton in the vector meson production process remains
intact, we suggest describing the t-dependence at the proton end by the proton
form factor $F_{1p}(t)$. In this way we arrive at  summation model S1:
\beq
{\mathcal P}_{S1}(s,t,Q^2) \ = \ 
{\mathcal P}_{np}(s,t,Q^2) \ + \ F_{1p}(t) \ {\mathcal P}_{p}(s,Q^2)~,
\label{eqn:rootsum_prop_related_expr}
\eeq
where ${\mathcal P}_{np}, \,  {\mathcal P}_{p}$ are given by
(\ref{eqn:offshell_nonpert_prop_related_expr}),
(\ref{eqn:offshell_pert_prop_related_expr}) respectively.


\subsubsection{Summation model S2}

Another possibility is to model the ``hard'' component with the CR
term ${\mathcal P}_{CR}$,(\ref{eqn:P_CR_general_form}).
The $t$-\dep\ is thus automatically provided, but the CR approach
does not specify the energy \dep.  In order to obtain a global
description of vector meson production, Regge energy \dep\ corresponding to 
the ``hard pomeron'' 
\cite{{DL98},{Donnachie:1999qv},{DL00}}
is introduced into the CR term by hand, in much same way that it was
introduced in the \nonpert\ term   using the soft \pom\ trajectory.
In this way one obtains summation model S2:
\beq
{\mathcal P}_{S2}(s,t,Q^2) \ = \ 
{\mathcal P}_{np}(s,t,Q^2) \ + \ {\mathcal P}_{CR}(s,t,Q^2)
\label{eqn:crsummodel_prop_related_expr}
\eeq
where ${\mathcal P}_{np}, \, {\mathcal P}_{CR}$ are given by 
(\ref{eqn:offshell_nonpert_prop_related_expr}),(\ref{eqn:P_CR_general_form})
respectively.

\section{Vector Meson Wave Functions}

The choice of the vertex function is crucial in  vector-meson
production models as it determines the virtualities dominating the integral
over the \qql;  the overall
normalisation and  \qsq\  \dep\ of the \cs; and  the
longitudinal to transverse ratios.
Unfortunately the detailed forms of the 
vertex functions  are unknown and only their general
analytical properties are established from various constraints
\cite{Zhitnitsky:1996}.  Therefore in practice, the chosen vertex functions 
provide an
essentially phenomenological description of the valence quark content
of the vector meson.
Here we shall consider possible forms for the vertex functions  by
starting from phenomenological wave functions for vector mesons
in their rest frame; and then boosting to the light-cone taking into account
the off-shell nature of the quark line.

\subsection{Wave Functions in Centre-of-Mass Frame}
\label{subsection:Wave_Functions_in_the_Centre-of-Mass_Frame}

The most popular choice  \cite{{Cudell_Royen:1998},{Schlumpf:1994},
{Huang:1994},
{Chibisov_Zhitnitsky:1995},{Krutov_Troitsky:1997},{de_Melo:1997},
{Belyaev_Johnson:1997}} of vector-meson
\wfs\ is suggested by long distance physics. This tells
us that a hadron at rest can be described to a good approximation as a 
system of constituent
quarks moving in a harmonic oscillator potential with a Gaussian \wf\
\beq
\Phi_{G} \! \left( \bfLsq  \right) \ = \ 
{\mathcal N}_{G} \exp \left(- \frac{\bfLsq }{2 p_F^2} \right)
\label{eqn:Gaussian_WF_dep_on_Lsq}
\eeq
where \bfLsq\ is the squared 3-momentum of either the quark or anti-quark, 
$p_F$ is the Fermi momentum and $\mathcal N$ is the normalisation.
We  investigated  five alternatives, the details of which can be found in
\cite{JG_thesis:2000}. The first is the power-law \wf\
\cite{{Schlumpf:1994},{Krutov_Troitsky:1997}}
\beq
\Phi_{pl}  \! \left( \bfLsq  \right) \ = \ 
{\mathcal N}_{pl} \left( 1 + \frac{\bfLsq}{\tilde{p}_F^2} \right)^{-n}
\label{eqn:Power-Law_WF_dep_on_Lsq}
\eeq
with $n = 2$ in our case. This was found not to be an acceptable choice as
it was not possible to obtain even a qualitative description of the data,
particularly for the longitudinal/transverse ratio for the $\rho$ and $\phi$
which are rather sensitive to the wave function details.

The four other \wfs\ are obtained by solving
the \nonrel\ Schroedinger equation with four different potentials
\cite{the_four_wfs:1995}. The first three of these are: 

$\bullet$ a power-law potential \cite{power_law_potential:1980} 
\beq
- a_1^2 \ + \ a_2^2 \left( \frac{r}{r_0} \right)^{0.1}
\label{eqn:VMWF_nonrel_power_law_pot}
\eeq
$\bullet$ a logarithmic potential \cite{log_potential:1977_1979}
\beq
b_1^2 \ + \ b_2^2 \log \left( \frac{r}{r_0} \right)
\label{eqn:VMWF_nonrel_logarithmic_pot}
\eeq
$\bullet$ a Coulomb-plus-linear potential (the Cornell potential)
\cite{coulomb_plus_lin_potential:1978}
\beq
- \frac{c_1^2}{r} \ + \ \frac{r}{c_2^2} \ + \ c_3^2
\label{eqn:VMWF_nonrel_Coulomb_plus_linear_pot}
\eeq
where the $a_i,b_i,c_i$ are various model-dependent parameters. 
The fourth is the
QCD-inspired potential of Buchm\"{u}ller and Tye
\cite{buchmuller_tye:1981}, which has a rather complicated 
position-space form.  It is linear at large distances and
quasi-Coulombic at short distances.  The deviations from pure
Coulombic behaviour reproduce the running of the strong coupling
constant, and the global shape of the potential is essentially
determined by two parameters --- the QCD scale $\Lambda$ and the QCD
string tension motivated by the light meson data. 
Non-relativistic \wfs\  are  reliable only
if the meson and both constituent quarks are heavy compared to the
average internal momentum,
and it is still not  clear whether one should use them
for the $J/\Psi$.  In particular Frankfurt, Koepf and 
Strikman\cite{Frankfurt_Koepf_Strikman:1997} have used  \wfs\ from 
various non-relativistic potential models to show that the integration region
where the quark's transverse momentum is larger than the charm mass can
contribute up to one third of the \qql\ integral in  \jpsi\
production.
For the \r\ and \f\ such  \wfs\ can be still be  considered  as  
an alternative choice, albeit without any firm foundation.

Finally, all these \vmwfs\ have to be normalised to reproduce the leptonic
decay width of the meson $\Gamma_{V \to e^+e^-}$ in the \vm\ rest frame.  
Furthermore, the normalisation  has
to be calculated for one quark leg off-shell to reflect the
kinematics of the \vmp\ models of
\fig\ref{CR_advanced_model:main_diagram}.  The derivation of the 
normalisation, and its reduction to the on-shell case in the appropriate limit,
has been given by Cudell and Royen\cite{Cudell_Royen:1998}. Their final result is 
\beq
\Gamma \ = \ 
\frac{64 \left( \aEM {\mathrm e}_q \right)^2}{9 \pi M_V^3}
\left[ \ 
i \pi \ {\mathcal G} \! \Big( \modbfL_d \Big)
\ + \
P \! \int_0^\infty \!
\frac{{\mathcal G} \! \Big( \modbfL \Big)}{\modbfL - \modbfL_d}
\ d \modbfL
\ \right]^2
\label{eqn:CR_offshell_VMWF_decay_width_final}
\eeq
where
\beq
{\mathcal G} \! \Big( \modbfL \Big) \ = \ 
\frac{\bfLsq}{\sqrt{\bfLsq + \mqsq}} \
\frac{2 \bfLsq + 3 \mqsq}
{\modbfL + \modbfL_d} \
\PhiLsq  \; ,
\label{eqn:CR_offshell_VMWF_function_G}
\eeq
\beq
\modbfL_d \ = \ \sqrt{ \quarter M_V^2 - \mqsq}~.
\label{eqn:CR_offshell_VMWF_Lsq_pole}
\eeq
and $m_q$ is the constituent quark mass. For $u,\,d$ quarks we take 
$m_q = 0.3$ GeV, for strange quarks $m_s = 0.45$ GeV and for charmed quarks
$m_c = 1.5$ GeV. 

The next step is to transform the \vm\ \cm\ wavefunctions into the
vertex functions $\Phi \! \left(z,  | \tildeellt |  \right)$ 
used in  (\ref{eqn:CR_offshell_final_amplitude}). This is done 
by first rewriting  the centre-of-mass wavefunctions in terms 
of invariants, which are then re-expressed in terms of  the \lc\ 
variables $z$ and $\tildeellt$ to obtain the vertex function.
We do this first using the Cudell-Royon prescription \cite{Cudell_Royen:1998}
for the quark or antiquark off-shell, and then show that if both are put 
on-shell it goes over to the Brodsky-Lepage prescription 
\cite{brodsky_etal_light_cone_pert_theory:1980}.


\subsection{The Cudell-Royon prescription}
\label{sect:boost_to_the_LC}

In \fig\ref{CR_advanced_model:main_diagram} the quark
and the meson have four-momenta $P_q = \left[ \frac{1}{2} V + \ell
\right]$ and $V$ respectively.
In the meson's \cm\ frame,  $P_q^\mu = \left( E_q, \ \bfL
\right)$ and  $V^\mu = \left( M_V, \ {\mathbf 0} \right)$ so that the
invariant quantity $P_q^\mu V_\mu = \left[ \frac{1}{2} V^\mu 
+ \ell^\mu \right] V_\mu$  becomes
$M_V E_q$. This gives
%
\beq
E_q \ = \ \frac{1}{M_V}  \left[ \half V + \ell \right] . V
\eeq
and
\beq
\bfLsq = E_q^2 \ - \ P_q^\mu P^{ }_{q \, \mu} 
\ = \
\left[ \frac{\ell . V}{M_V} \right]^2 \ - \ \ell . \ell
\label{eqn:CR_prescription}
\eeq
for the squared three momentum of the quark in the centre of mass frame, 
where no assumption about the on-shell or off-shell nature  of the quark 
has been made.
Thus one obtains the relation 
\beq
\Phi_{C \! M} \left( \bfLsq \right) \, \longleftrightarrow \, 
\Phi_{LC} \left(
\left[ \frac{\ell . V}{M_V} \right]^2 - \ell . \ell 
\right).
\label{eqn:CR_prescription_in_WF}
\eeq
between the \vmwf\ $\Phi_{C \! M}$ expressed
in the \cm\ variable $\bfLsq$ and $\Phi_{LC}$ expressed in terms of
appropriate invariants.
This is the Cudell-Royen(CR) prescription \cite{Cudell_Royen:1998}.
One finally  has to rewrite the four-vector $\ell$ in terms of the 
variables  $z$ and $\tildeellt$, using equations given by 
Cudell and Royen\cite{Cudell_Royen:1998}\footnote{ In
\cite{Cudell_Royen:1998} our $\tildeellt$ is denoted by $l_t$. }
which take account of 
the fact that the quark is off-shell and the anti-quark is on-shell
\beq
\left[ \half V - \ell \right]^2  =  \mqsq
\hspace{1cm}
\left[ \half V + \ell \right]^2  \ne  \mqsq
\label{eqn:CR_prescr_antiquark_onshell}
\eeq
in the two diagrams of \fig\ref{CR_advanced_model:main_diagram}.
The  asymmetry between the  quark and antiquark results in an
asymmetry under the transformation $ z  \leftrightarrow
1-z $, which interchanges the ``+'' momentum carried by the quark and 
antiquark.  
However,  when calculating vector-meson production, one must also take into
account diagrams corresponding to those in
\fig\ref{CR_advanced_model:main_diagram} but with the quark charge
flow reversed. The cut conditions then put the quark
(instead of the anti-quark) on-shell, giving the reverse of
\eqn(\ref{eqn:CR_prescr_antiquark_onshell}).
Thus the asymmetry present within each pair of diagrams is no longer
present once the vector-meson production amplitudes from all four diagrams 
are summed.

\begin{figure}[!p]
\vspace*{-5mm}
\begin{minipage}[t]{0.49\textwidth}
\begin{center}
\psfig{file=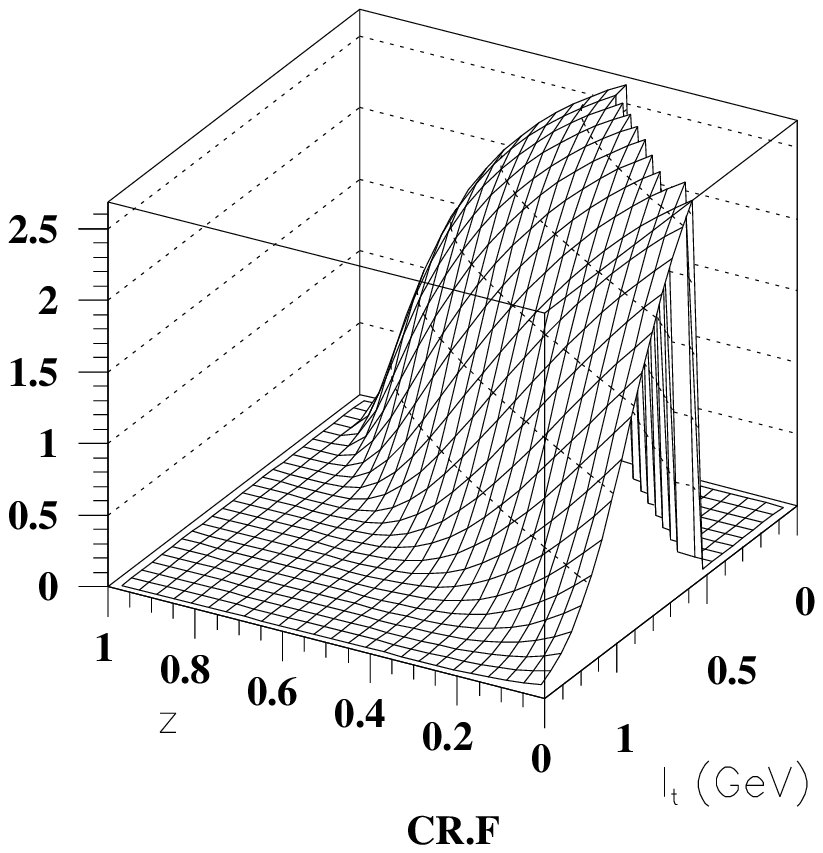, width=\textwidth}
\vspace*{-10mm}
\caption{ 
\label{cr1}
\rmes\ relativistic Gaussian \wf\ (CR precription) with $p_F = 0.6 \, \gev$, 
with ${\mathcal N} = 2.803. $
}
\end{center}
\end{minipage}
\hfill
\begin{minipage}[t]{0.49\textwidth}
\begin{center}
\psfig{file=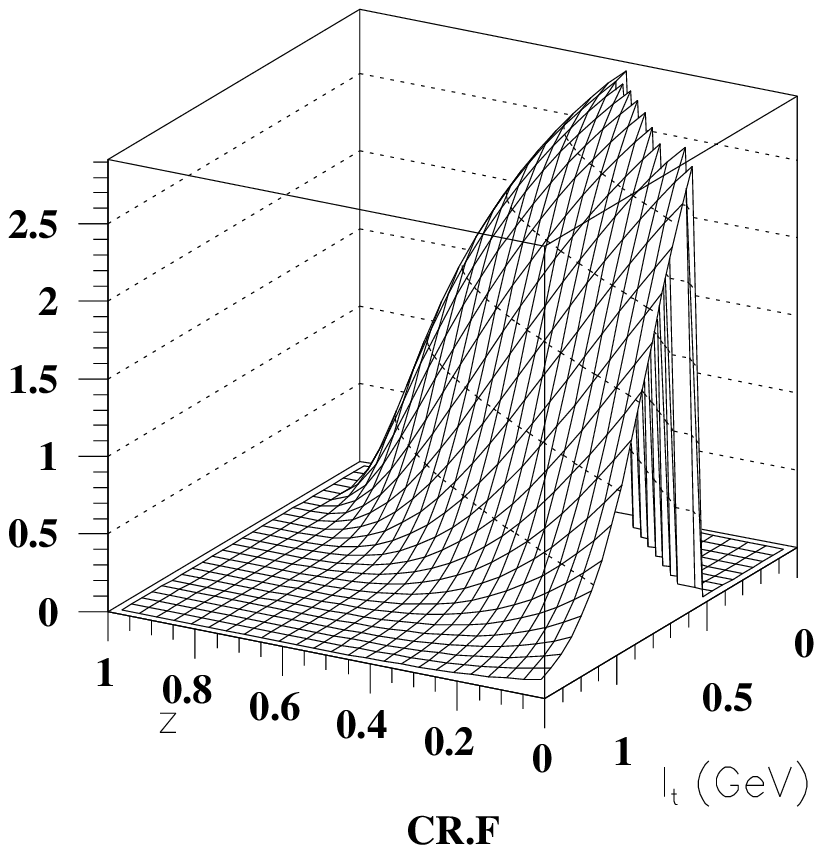, width=\textwidth}
\vspace*{-10mm}
\caption{ 
\label{cr2}
\rmes\ non-relativistic logarithmic potential \wf (CR precription)
and ${\mathcal N} = 2.763 $.
}
\end{center}
\end{minipage}
\end{figure}

As an illustration, the vertex functions $\Phi(z,\left| \bfellt \right|)$  
corresponding to 
the Gaussian and logarithmic potential \vm\ \cm\ \wfs\ 
(\ref{eqn:Gaussian_WF_dep_on_Lsq})
are shown in \figs\ \ref{cr1} and \ref{cr2}, 
where the  asymmetry under  $ z  \leftrightarrow
1-z $ is clearly seen. 
One also sees that the vertex functions
are centred towards the point $z=0$,
$|\bfellt| = 0$, implying that the most likely configurations
are those where the off-shell quark carries only a small fraction 
$zq$ of the meson's longitudinal momentum, while the on-shell
anti-quark carries most of it, namely $(1-z)q$.
However the vertex functions  never actually reach the 
point $z=0$, $|\bfellt|=0$,
because the region around it is unphysical, corresponding to
$\bfLsq < 0$. The border of this region  
correspond to $\bfLsq = 0$, where  the
quark and  anti-quark have no relative momentum in the meson rest
frame. 
In the other case, with the anti-quark off-shell
and the quark on-shell, the CR prescription  gives similar
vertex functions, but centred towards $z=1$ and $| \bfellt | = 0$
instead of $z=0$ and $| \bfellt | = 0$ so that the on-shell particle 
again carries most of the meson's
longitudinal momentum.

%
%

In the numerical calculations  we focussed mainly\footnote{A  fuller
discussion of the various possible wavefunctions is given in
\cite{JG_thesis:2000}.} on Gaussian
wavefunctions with 
values of $p_F$ ranging from 0.1 \gev\ to
0.6 \gev\ for \r\ and \f\ mesons and from 0.2 \gev\ to 1.2 \gev\ for
the \jpsi. The motivation behind  these ranges of
values 
is the transverse size of the corresponding \vm; in particular, that for 
the light vector mesons is similar to that used in the literature
for the pion
\cite{{Schlumpf:1994},{Huang:1994},{Krutov_Troitsky:1997}}. As we shall
see, our best results for electroproduction were obtained with
$p_F = $ 0.5,0.6,1.2  \gev\ for the $\rho$, \f\ and \jpsi\ mesons, 
for which the appropriate values of the normalisation
constant are ${\mathcal N}_{G} = $ 3.597, 3.226 and 2.905. 
  

\subsection{Relation to the Brodsky-Lepage prescription}

The prescription of Brodsky and Lepage \cite{brodsky_etal_light_cone_pert_theory:1980},
often found in the literature
\cite{{Huang:1994},{Frankfurt_Koepf_Strikman:1997},
{Nemchik_Nikolaev_Zakharov_Predazzi:1997}},
connects  the \wfs\
in the \cm\ frame and the \lc\ frame by equating the off-shell
propagator $\varepsilon = \left[ M^2 - \left( \sum_{i=1}^n k_i
\right)^2 \right]^{-1}$ in the two frames.
In the general case the propagator for a particle of mass $M$ whose
constituents $i = 1, \ldots , n$ have masses $m_i$ 
is given by
\beq
\varepsilon^{-1} \ = \ 
\left\{ 
\begin{array}{lllr}
M^{2} - \Big( \sum^{n}_{i=1} L^{0}_{i} \Big)^{2} \quad\ &
\sum^{n}_{i=1} {\mathbf L}_{i} = 0 \quad\ & \quad\ & 
[CM] \\
M^{2} - \sum^{n}_{i=1} 
\left( \bfelltsq + m_i^2 \right) / z_i \quad\ &
\sum_{i=1}^{n} \bfell_{t i} = 0 \quad\ &
\sum_{i=1}^{n} z_i = 1 \quad\ &
[LC] \\
\end{array}  \right.
\label{eqn:BHL_prescription_general}
\eeq
where $L_i^\mu = \left( L_i^0, {\mathbf
L}_i \right)$ and $z_i$, $\bfell_{t i}$ are the centre of mass four momenta 
and \lc\ variables respectively.
For a system of two-particles of equal masses this gives $\bfL_1 =
-\bfL_2$.
The relation $L_1^0 = L_2^0$ is then used \cite{Huang:1994} 
implying that both constituent particles are
on-shell.

According to the Brodsky-Lepage(BL) prescription the propagators in 
both frames are equated yielding
\beq
\bfLsq \ = \ 
\frac{ \bfelltsq + \mqsq }{4 z \left( 1 - z \right)} - \mqsq
\label{eqn:BHL_prescription}
\eeq
and
\beq
\Phi_{C \! M} \left( \bfLsq \right) \, \longleftrightarrow \,
\Phi_{LC} \left(
\frac{ \bfelltsq + \mqsq }{4 z \left( 1 - z \right)} - \mqsq
\right)~.
\label{eqn:BHL_prescription_in_WF}
\eeq
Alternatively, one could  equate the expressions for invariant
mass of the \qqp\ in the \vm\ \cm\ and \lc\ frames
\beq
M_X^2 \ = \ 4 \left( \bfLsq + \mqsq \right) = \frac{\bfelltsq + 
\mqsq}{z \left( 1 - z \right)}
\label{eqn:BHL_qq_inv_mass_in_CM}
\eeq
yielding the same prescription (\ref{eqn:BHL_prescription}). The functions  
(\ref{eqn:BHL_prescription_in_WF}) are often called \lcwfs; and for 
quarks and antiquarks of equal mass, the prescription is seen to be 
explicitly  symmetric under  $ z  \leftrightarrow 1-z $.


The BL prescription is not applicable in the kinematical framework
of  \fig\ref{CR_advanced_model:main_diagram}, in which the quark (or
antiquark) is off-shell. 
On the other hand the CR prescription  (\ref{eqn:CR_prescription}), which
allows for this, reduces 
to the BL prescription if both both particles are put on-shell
\beq
\left[ \half V - \ell \right]^2  =  \mqsq
\hspace{1cm}
\left[ \half V + \ell \right]^2  =  \mqsq
\label{eqn:CR_prescr_both_quarks_onshell}
\eeq
yielding
\beq
\ell . V  =  0 
\hspace{1cm}
\ell . \ell  =  \mqsq - \quarter M_V^2 
\label{eqn:ell_dot_V_is_zero_etc}
\eeq
but not in general  $\ell . \ell = 0$ or $\ell^\mu \equiv
0$.
Substituting (\ref{eqn:ell_dot_V_is_zero_etc}) into
(\ref{eqn:CR_prescription}) gives
%
%
\beq
\bfLsq \ = \ \quarter M_V^2 - \mqsq~.
\label{eqn:Lsq_both_q_onshell}
\eeq
so that the square of the three-momentum of the quark in meson rest frame
is fixed by the mass of the meson $M_V$ and by the choice
of the mass of the quark $m_q$.
However, \bfLsq\ is not zero provided the $m_q$ is not chosen to
be exactly half of the meson mass. 
Identifying the mass of the \vm\ $M_V$ with the invariant mass $M_X$
of the \qqpair\ and substituting
(\ref{eqn:BHL_qq_inv_mass_in_CM}) into (\ref{eqn:Lsq_both_q_onshell}),
one again obtains the BL prescription (\ref{eqn:BHL_prescription}).

If the quark mass is exactly
half of the meson mass, 
Eq.(\ref{eqn:ell_dot_V_is_zero_etc}) gives
\beq
\ell . V  =  0 
\hspace{1cm}
\ell . \ell   =  0
\label{eqn:ell_is_zero}
\eeq
corresponding to a \qqp\ with zero relative momentum in the meson rest frame.
The quark and the anti-quark  share equally the meson's four-momentum.  In
the \lc\ variables it reads
\beq
\delta^{(2)} \left( \bfellt \right) \
\delta \left( z - \frac{1}{2} \right)
\label{eqn:CR_prescr_simplest_WF}
\eeq
which  has also been used as a very basic \lcwf\ in the literature
\cite{{DL:1987},{Cudell_Royen:1996},{horgan_etal:1982}}.

\begin{figure}[!p]
\vspace*{-5mm}
\begin{minipage}[t]{0.49\textwidth}
\begin{center}
\psfig{file=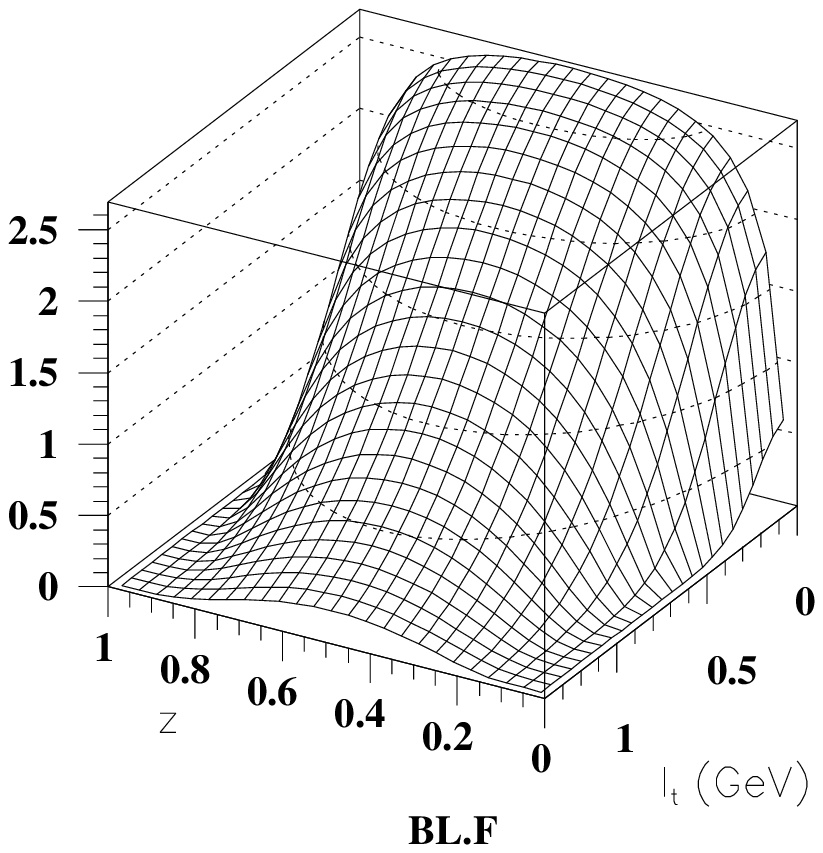, width=\textwidth}
\vspace*{-10mm}
\caption{ 
\label{bl1}
\rmes\ relativistic Gaussian \wf\ $p_F = 0.6 \, \gev$, BL precription.
}
\end{center}
\end{minipage}
\hfill
\begin{minipage}[t]{0.49\textwidth}
\begin{center}
\psfig{file=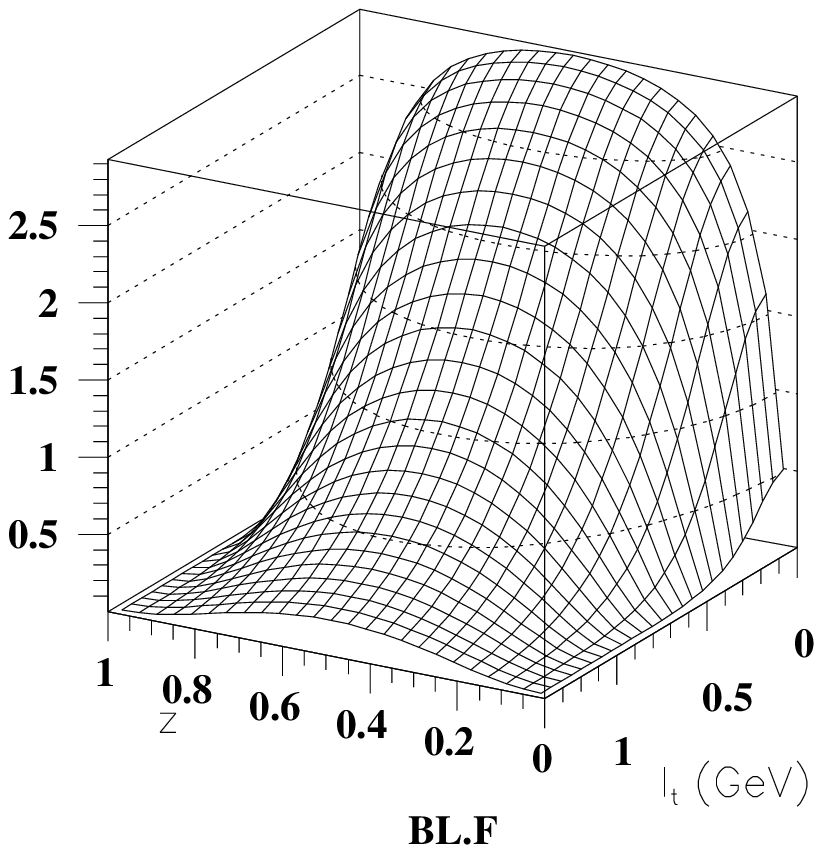, width=\textwidth}
\vspace*{-10mm}
\caption{ 
\label{bl2}
\rmes\ non-relativistic logarithmic potential \wf, BL prescription.
}
\end{center}
\end{minipage}
\end{figure}

Figures \ref{bl1} and \ref{bl2} show  vertex functions obtained
using the Brodsky-Lepage prescription(i.e. \lc\ wavefunctions), to be 
compared with the vertex
functions of Figures \ref{cr1} and \ref{cr2} which were obtained 
using the Cudell-Royon prescription from same centre of mass wavefunctions.
As can be seen, Brodsky-Lepage wave functions  are symmetric 
under  $ z  \leftrightarrow 1-z $ 
with a maximum at $z =\frac{1}{2}$, $|\bfellt| = 0$. 
In other words the most probable configuration is  where
the quark and antiquark  share equally the vector meson's longitudinal
momentum.
Further, only one single value(``height'') of the \wf, determined by
the value of $\bfLsq$, enters in any given \vmp\ calculation since the
$\bfLsq$ is fixed by (\ref{eqn:Lsq_both_q_onshell}) once 
the masses of the quarks and the vector meson are fixed.
In calculating (\ref{eqn:CR_offshell_final_amplitude}), this can be
imposed via a separate condition stating the on-shellness of
the quarks.  Such a condition, reducing the dimensions of integration,
is indeed present in the vector-meson production models exploiting 
the BL prescription \cite{{Cudell_Royen:1996},{DL:1987}}.


%
%

\section{Results}
We have already said that the power-law wave function 
(\ref{eqn:Power-Law_WF_dep_on_Lsq}) is not appropriate, and we do not 
discuss it further. The results obtained using the wave functions obtained 
from the power-law potential (\ref{eqn:VMWF_nonrel_power_law_pot}), the
logarithmic potential (\ref{eqn:VMWF_nonrel_logarithmic_pot}), the
Coulomb-plus-linear potential (\ref{eqn:VMWF_nonrel_Coulomb_plus_linear_pot})
and the Buchm{\" u}ller-Tye potential \cite{buchmuller_tye:1981} are almost 
identical. They
predict successfully the longitudinal/transverse ratio for each of the
\r\, \f\ and $\jpsi$ and the correct shape for $d\sigma/dQ^2$ in each case.
However it is not possible to obtain simultaneously the correct normalisation 
of all the cross sections. If the normalisation is adjusted to fit the \r\
cross section, say, then the predicted $\jpsi$ cross section is too high.
Conversely, if the normalisation is adjusted to the $\jpsi$ cross section
then the predicted \r\ cross section is too low. The problem is that there
is no flexibility in the wave functions: they are all fixed by the
parameters of the potentials. This is not the case for the Gaussian
wave function (\ref{eqn:Gaussian_WF_dep_on_Lsq}) for which the parameter
$p_F$ can be adjusted independently for each case.

\begin{figure}[!p]
\vspace*{-5mm}
\begin{minipage}[t]{0.49\textwidth}
\begin{center}
\psfig{file=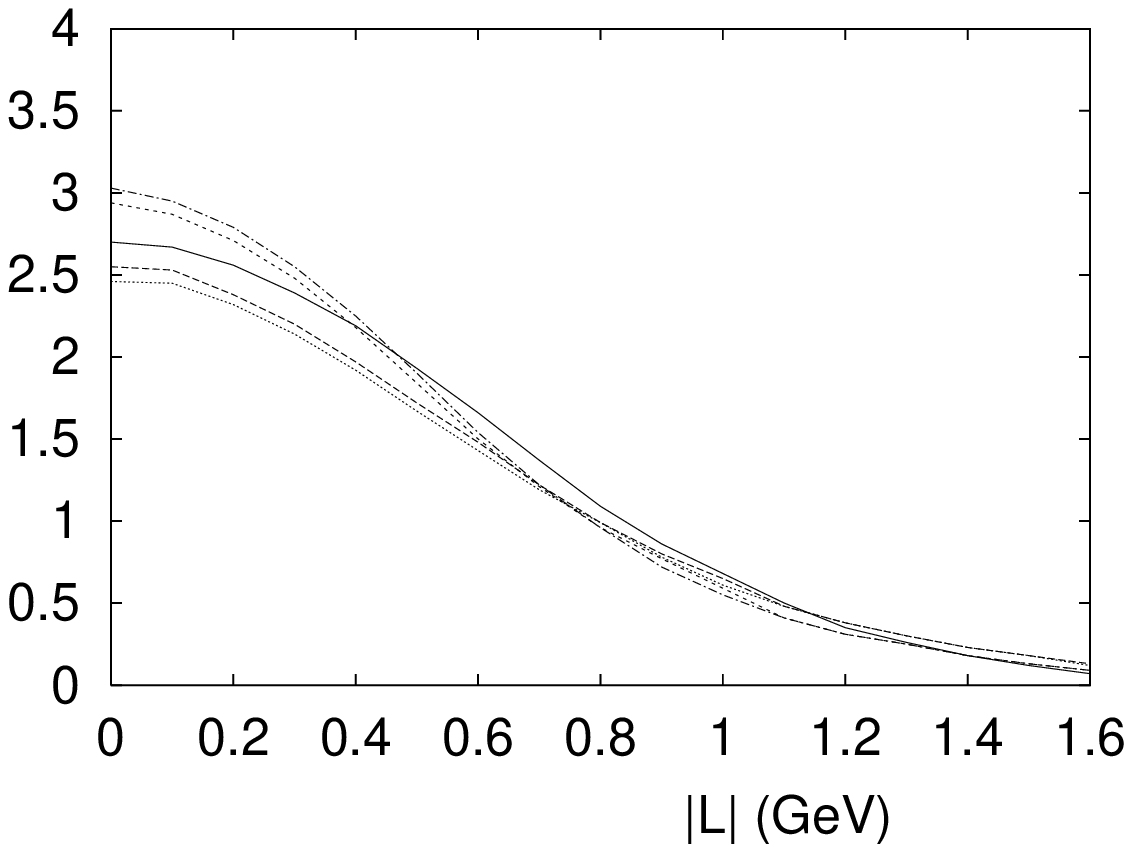, width=\textwidth}
\vspace*{-10mm}
\caption{ 
\label{rhowf}
Comparison of the Gaussian wave function (\ref{eqn:Gaussian_WF_dep_on_Lsq})
for the \r\ with $p_F = 0.5$ (solid line) with the wavefunctions obtained from
non-relativistic potentials. The dashed lines, from top to bottom,
correspond to the Buchm\"{u}ller-Tye, logarithmic, power-law and Coulomb
plus linear potentials respectively.
}
\end{center}
\end{minipage}
\hfill
\begin{minipage}[t]{0.49\textwidth}
\begin{center}
\psfig{file=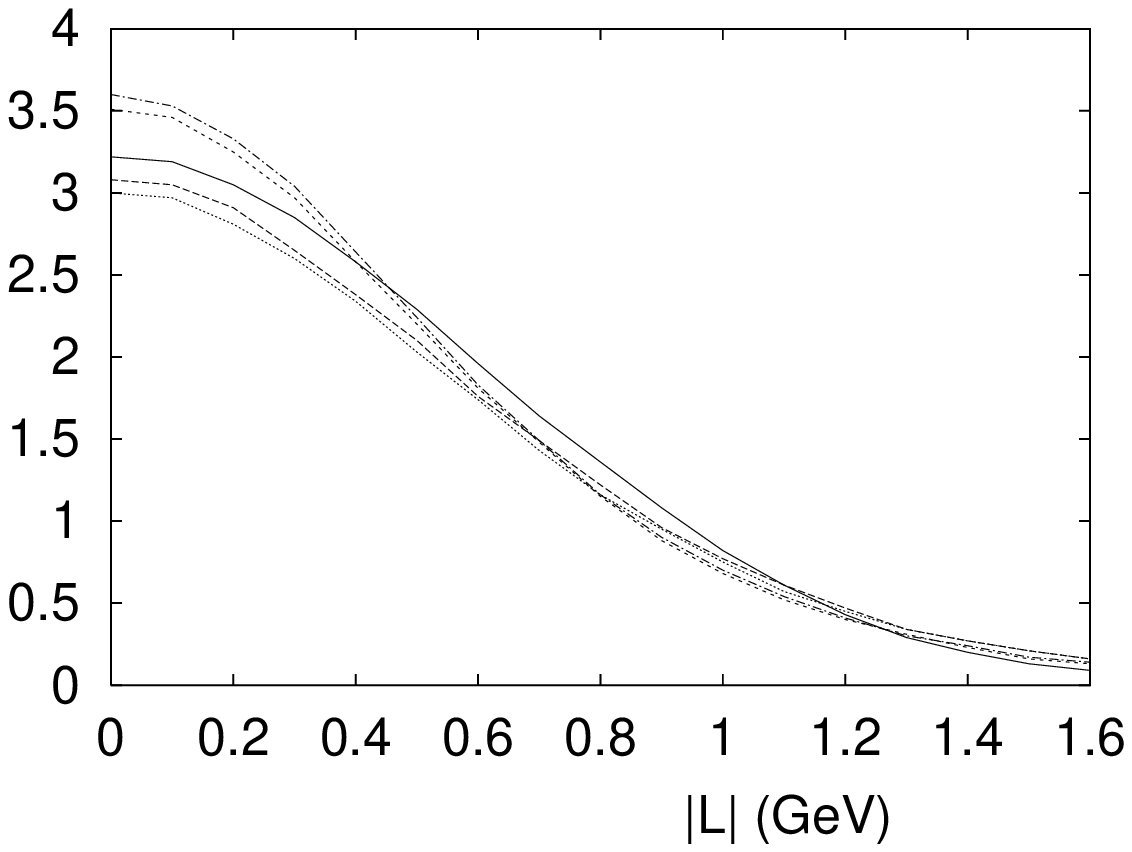, width=\textwidth}
\vspace*{-10mm}
\caption{ 
\label{phiwf}
Comparison of the Gaussian wave function (\ref{eqn:Gaussian_WF_dep_on_Lsq})
for the \f\ with $p_F = 0.6$ (solid line) with the wave functions obtained from
non-relativistic potentials.The dashed lines, from top to bottom,
correspond to the Buchm\"{u}ller-Tye, logarithmic, power-law and Coulomb
plus linear potentials respectively.
}
\end{center}
\end{minipage}
\end{figure}

\begin{figure}[!p]
\vspace*{-5mm}
\begin{minipage}[t]{0.49\textwidth}
\begin{center}
\psfig{file=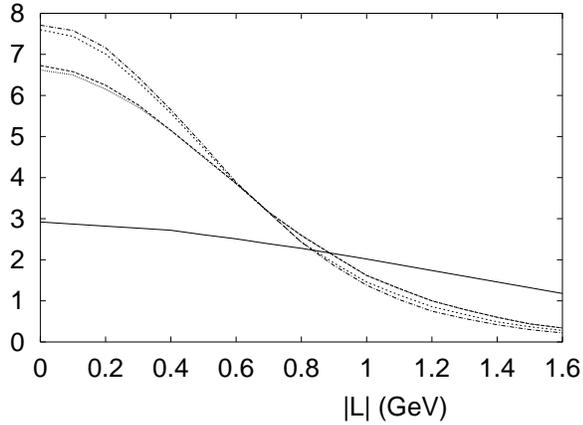, width=\textwidth}
\vspace*{-10mm}
\caption{ 
\label{psiwf}
Comparison of the Gaussian wave function (\ref{eqn:Gaussian_WF_dep_on_Lsq})
for the $\jpsi$  with $p_F = 1.2$ (solid line) with the wave functions 
obtained from
non-relativistic potentials. The dashed lines, from top to bottom,
correspond to the Buchm\"{u}ller-Tye, logarithmic, power-law and Coulomb
plus linear potentials respectively.
}
\end{center}
\end{minipage}
\hfill
\begin{minipage}[t]{0.49\textwidth}
\begin{center}
\end{center}
\end{minipage}
\end{figure}

It turns out that the Gaussian wave function corresponding to the best 
choice of $p_F$ is very close to the wave functions
obtained from the power-law, logarithmic, Coulomb-plus-linear and
the Buchm{\"u}ller-Tye potentials for the \r\ and \f\, but differs 
significantly
in the case of the $\jpsi$. This is shown in figures \ref{rhowf},
\ref{phiwf} and \ref{psiwf}, from
which it is obvious that for the $\jpsi$ the model requires a much narrower
transverse distribution in configuration space than is provided by the wave
functions obtained from solving the Schroedinger equation for specific
potentials.


For clarity of presentation we show results only for the Gaussian wave 
function, and comment on the principal differences in results obtained
with the alternative vertex functions. 
The parameters in our calculation are the $p_F$ for each of the $\rho$, \f\ 
and $\jpsi$, which control the 
transverse size of the vector-meson wave functions; the momentum cut-off
$Q_0^2$ when the gluon structure function is used in the perturbative term,
or the overall normalisation of the perturbative term in the Cudell-Royon 
model. The normalisation of the nonperturbative term is effectively fixed
by the $\rho$ photoprodution cross section which it dominates \cite{DL00}.
Two sets of results are shown: S1 refers to non-perturbative plus
gluon structure function; S2 refers to non-perturbative plus Cudell-Royon.
In both cases the same parameters for the two-gluon exchange contributions
are used for the $\rho$, $\phi$ and $\jpsi$, so that the only difference 
allowed for different vector mesons is in the value of the relevant $p_F$.

\begin{figure}[!p]
\vspace*{-5mm}
\begin{minipage}[t]{0.49\textwidth}
\begin{center}
\psfig{file=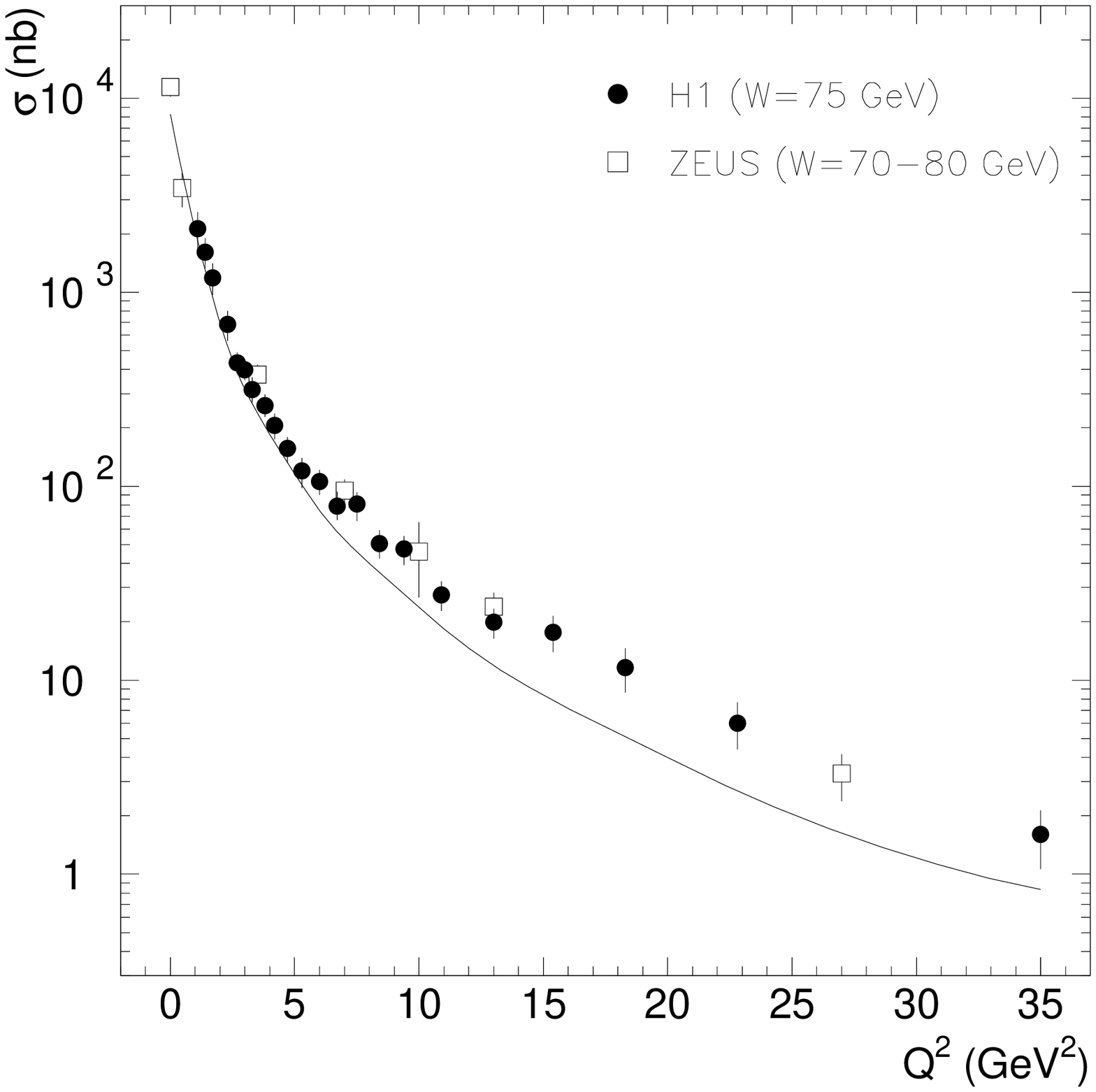, width=\textwidth}
\vspace*{-10mm}
\caption{ 
\label{S1r_a}
 $Q^2$ dependence of the \rmes\ cross-section at $W = 75 \, \gev$
in model S1.The data are from:  H1 \cite{H1_r_00};  and  ZEUS 
 \cite{ZEUS_r_95b}~\cite{ZEUS_r_98}~\cite{ZEUS_r_99}. 
}
\end{center}
\end{minipage}
\hfill
\begin{minipage}[t]{0.49\textwidth}
\begin{center}
\psfig{file=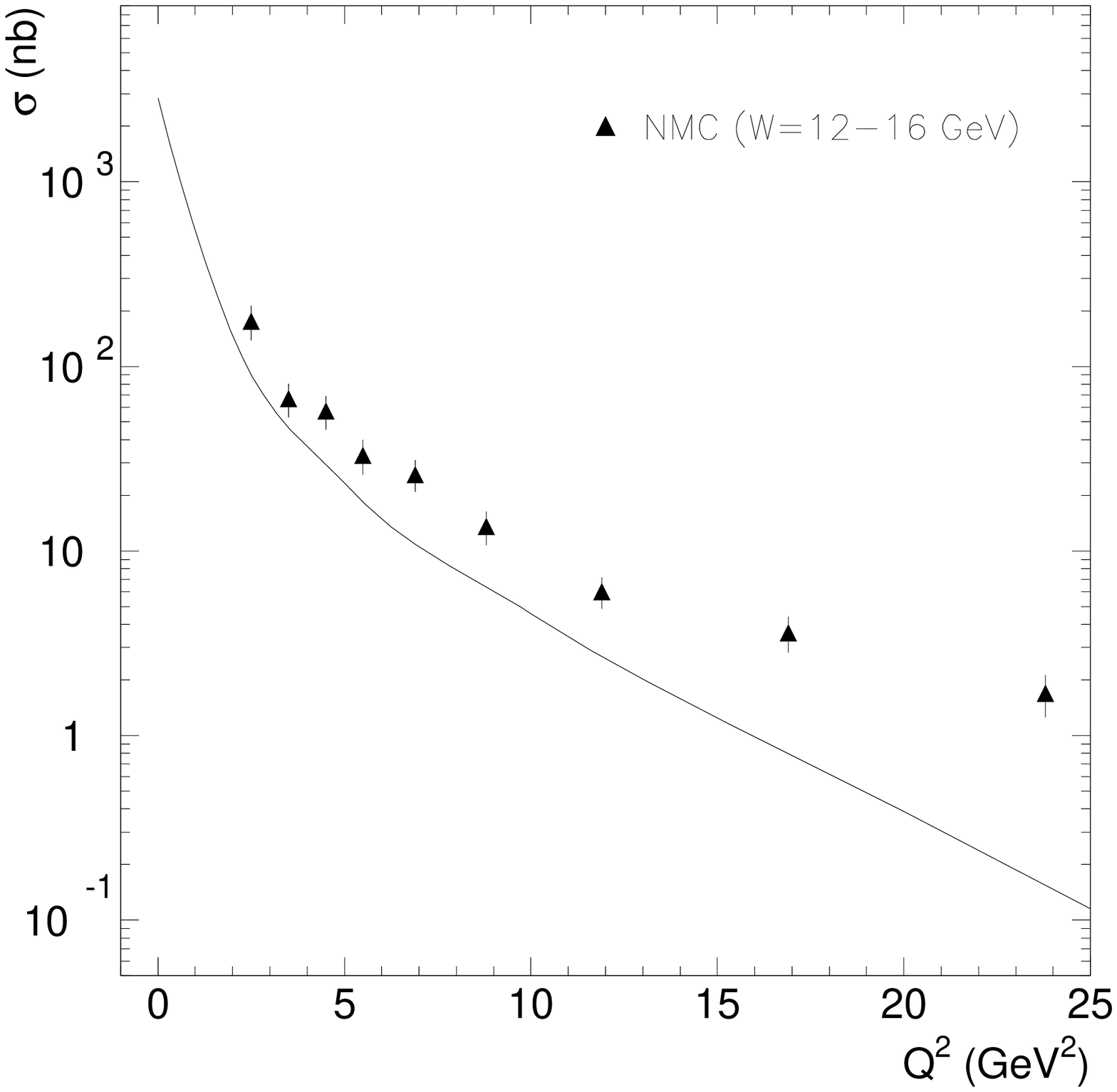, width=\textwidth}
\vspace*{-10mm}
\caption{ 
\label{S1r_b}
$Q^2$ dependence of the \rmes\ cross-section at $W = 15 \, \gev$
in model S1. The data are from: NMC  \cite{NMC_r_94}.
}
\end{center}
\end{minipage}
\end{figure}

\begin{figure}[!p]
\vspace*{-5mm}
\begin{minipage}[t]{0.49\textwidth}
\begin{center}
\psfig{file=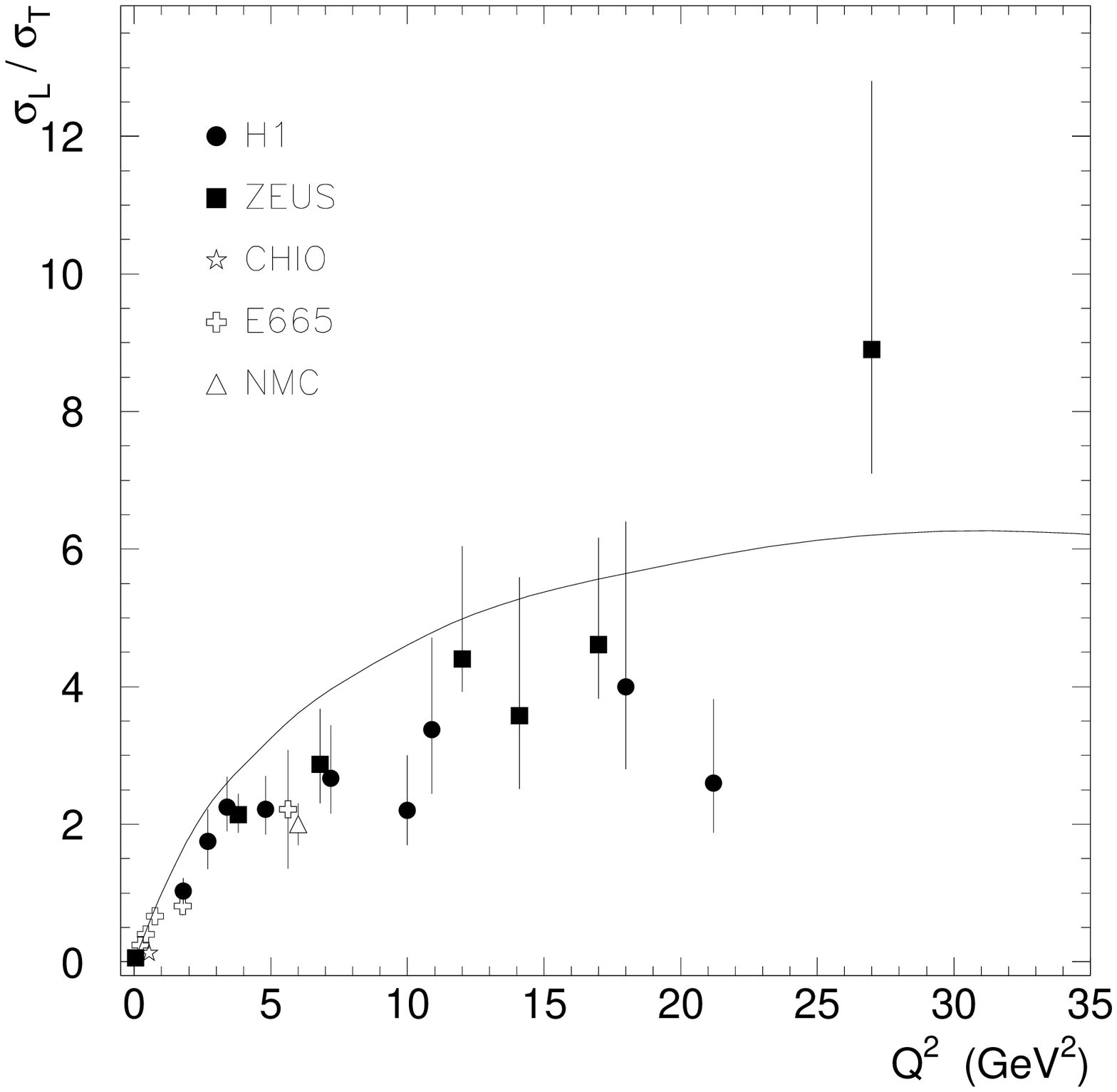, width=\textwidth}
\vspace*{-10mm}
\caption{ 
\label{S1r_c}
$Q^2$ dependence of the \rmes\  longtitudinal to transverse cross-section
ratio  at $W = 75 \, \gev$ in model S1.
The data are from:  CHIO \cite{CHIO_r_82};
 NMC  \cite{NMC_r_94}; E665  \cite{E665_r_97}; 
 H1  \cite{H1_r_96a}~\cite{H1_r_96b}~\cite{H1_r_00}; and
 ZEUS  \cite{ZEUS_r_95a}~\cite{ZEUS_r_95b}~\cite{ZEUS_r_99}. 
}
\end{center}
\end{minipage}
\hfill
\begin{minipage}[t]{0.49\textwidth}
\begin{center}
%
%
%
%
\end{center}
\end{minipage}
\end{figure}

\begin{figure}[!p]
\vspace*{-5mm}
\begin{minipage}[t]{0.49\textwidth}
\begin{center}
\psfig{file=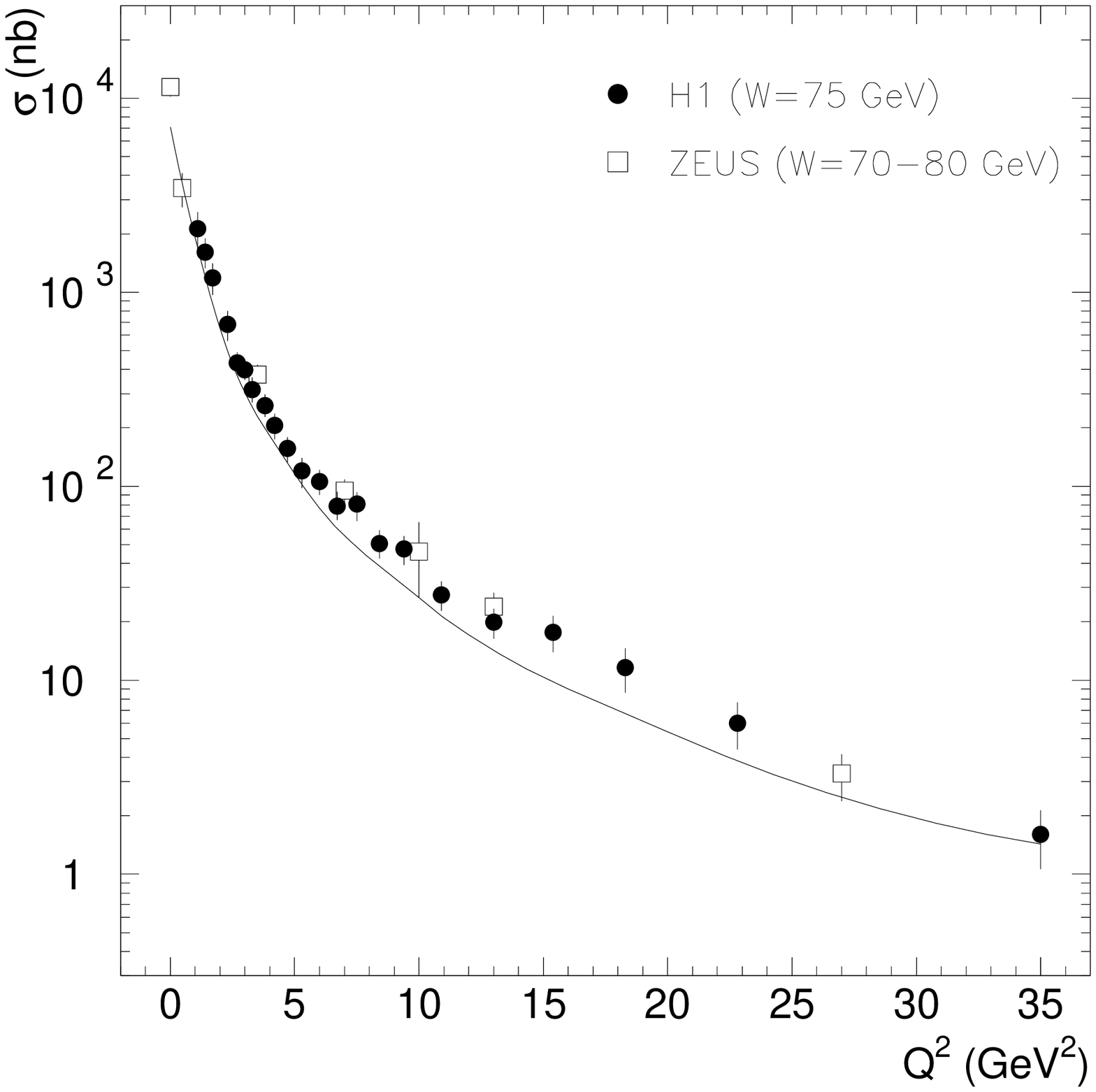, width=\textwidth}
\vspace*{-10mm}
\caption{ 
\label{S2r_a}
$Q^2$ dependence of the \rmes\ cross-section at $W = 75 \, \gev$
in model S2. The data are from:  H1 \cite{H1_r_00};  and  ZEUS 
 \cite{ZEUS_r_95b}~\cite{ZEUS_r_98}~\cite{ZEUS_r_99}. 
}
\end{center}
\end{minipage}
\hfill
\begin{minipage}[t]{0.49\textwidth}
\begin{center}
\psfig{file=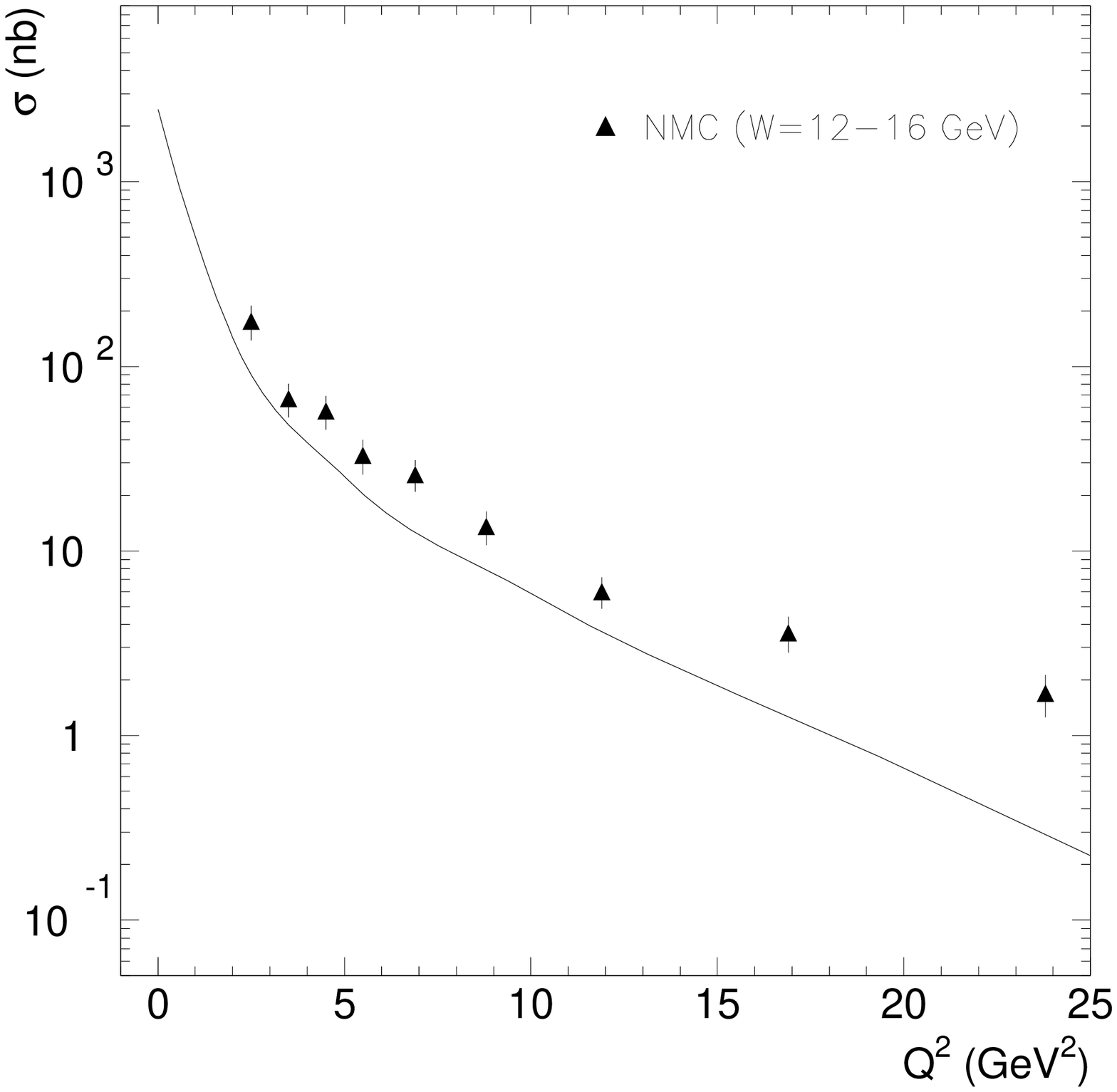, width=\textwidth}
\vspace*{-10mm}
\caption{ 
\label{S2r_b}
$Q^2$ dependence of the \rmes\ cross-section at $W = 15 \, \gev$
in model S2. The data are from: NMC  \cite{NMC_r_94}.
}
\end{center}
\end{minipage}
\end{figure}

\begin{figure}[!p]
\vspace*{-5mm}
\begin{minipage}[t]{0.49\textwidth}
\begin{center}
\psfig{file=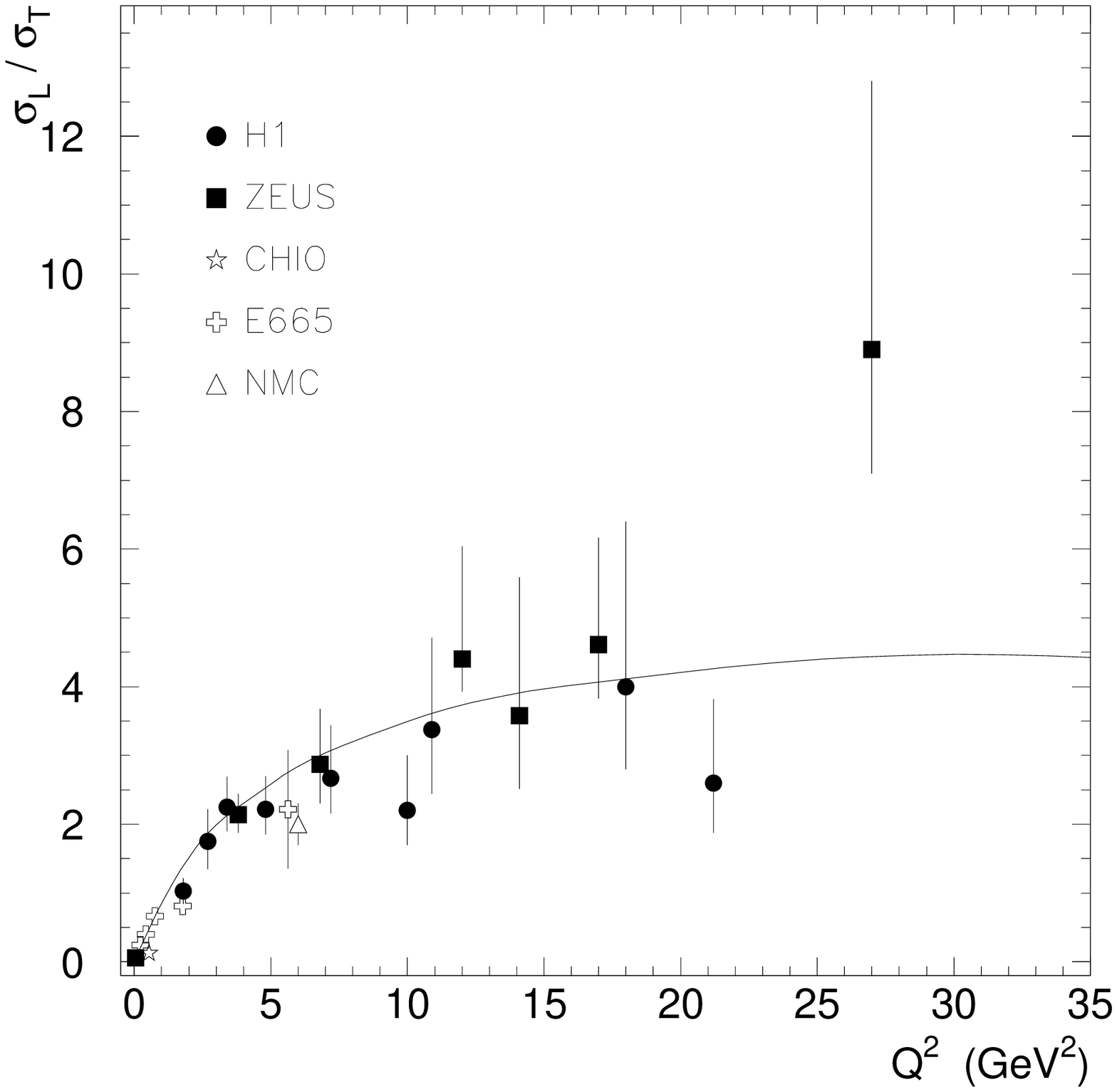, width=\textwidth}
\vspace*{-10mm}
\caption{ 
\label{S2r_c}
$Q^2$ dependence of the \rmes\  longtitudinal to transverse 
cross-section ratio  
at $W = 75 \, \gev$ in model S2. The data are from:  CHIO \cite{CHIO_r_82};
 NMC  \cite{NMC_r_94}; E665  \cite{E665_r_97}; 
 H1  \cite{H1_r_96a}~\cite{H1_r_96b}~\cite{H1_r_00}; and
 ZEUS  \cite{ZEUS_r_95a}~\cite{ZEUS_r_95b}~\cite{ZEUS_r_99}. 
}
\end{center}
\end{minipage}
\hfill
\begin{minipage}[t]{0.49\textwidth}
\begin{center}
%
%
%
\end{center}
\end{minipage}
\end{figure}

We start with the $\rho$. The results for $d\sigma/dQ^2$ for and 
$\sigma_L/\sigma_T$ with $p_F = 0.5$ GeV are shown in figures \ref{S1r_a}, 
\ref{S1r_b} and \ref{S1r_c} for S1 and in figures \ref{S2r_a}, \ref{S2r_b} and 
\ref{S2r_c} for S2. The result for $d\sigma/dQ^2$ at $W=75$ and for
$\sigma_L/\sigma_T$, which has little energy dependence, are satisfactory,
but that for $d\sigma/dQ^2$ at $W=15$ clearly falls below the data at all
$Q^2$. This is not surprising as at this lower energy there is a contribution
from reggeon exchange which we have not taken into account. The results for
$d\sigma/dQ^2$, other than the normalisation, are not strongly dependent on 
the choice of $p_F$. The shape is reproduced satisfactorily for $0.3 \le p_F
\le 0.6$ GeV. However $\sigma_L/\sigma_T$ depends very strongly on $p_F$, 
the ratio rising rapidly with decreasing $p_F$, and restricting the choice 
of $p_F$ to the upper end of the range. The S2 result is slightly better 
overall. The results obtained using the wave functions derived from the
specific potentials are very close to those shown, and provide an equally
satisfactory description. 

It is clear from figures \ref{S1r_a} and \ref{S2r_a} that an increase in 
normalisation would convert a good description of the data into an excellent 
one. If we were considering \r\ meson production in isolation then it would be 
appropriate to do this by increasing $\alpha^{(0)}_s$. However, as we shall 
see, this would then impact adversely on the cross sections for the \f\ and 
$\jpsi$, particularly on the former by making it too large. 
The problem of simultaneously obtaining the correct
normalisation for the photo-  and electroproduction of the \r\ and the \f\ 
within the constraints of pre-defined wave functions is well-known; for
example see \cite{DGP98}.

\begin{figure}[!p]
\vspace*{-5mm}
\begin{minipage}[t]{0.49\textwidth}
\begin{center}
\psfig{file=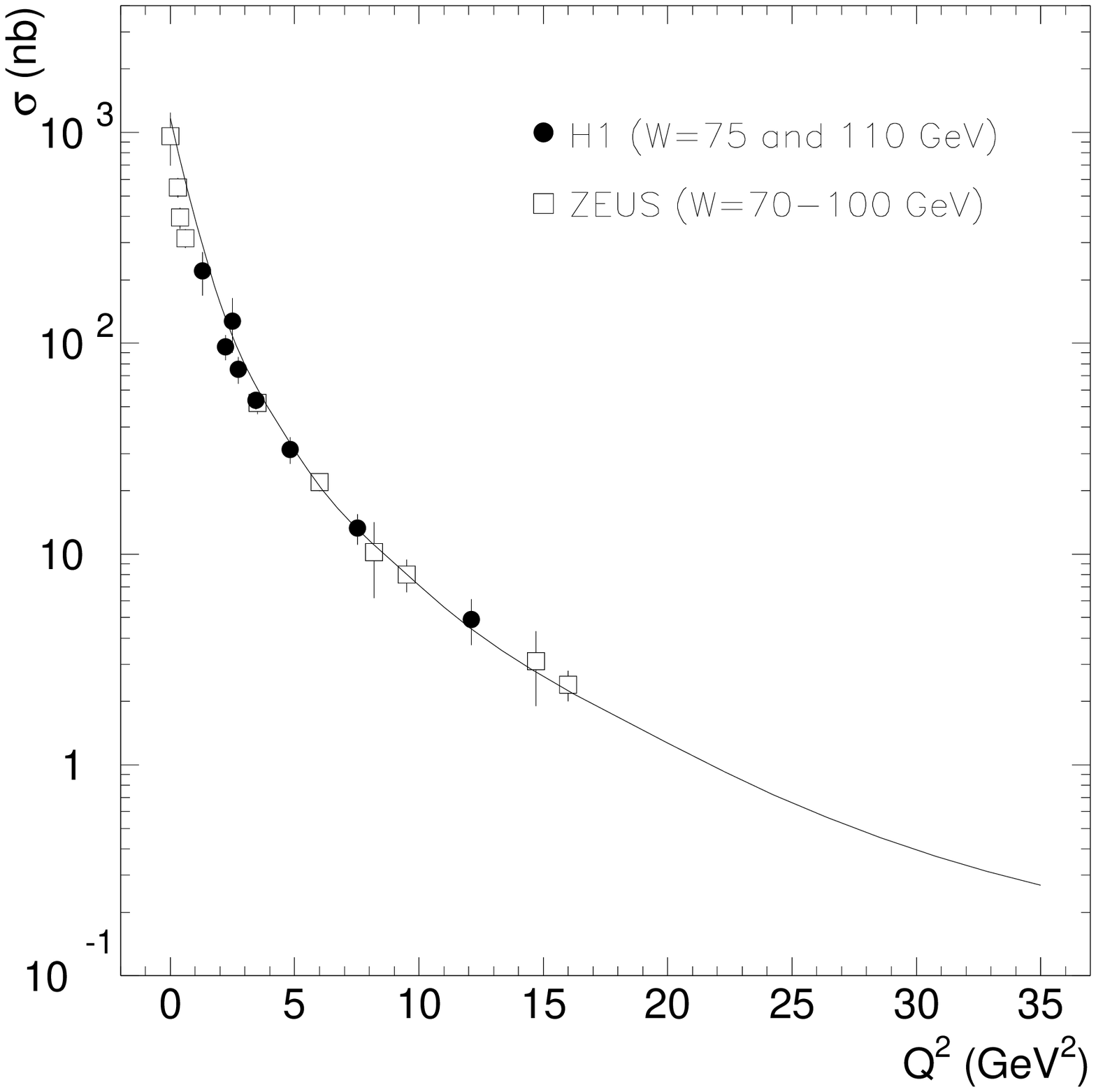, width=\textwidth}
\vspace*{-10mm}
\caption{ 
\label{S1p_a}
$Q^2$ dependence of the \fmes\ cross-section at $W = 90 \, \gev$ 
in model S1. The data are from: H1  \cite{H1_f_97}~\cite{H1_f_00};
and ZEUS  \cite{ZEUS_f_96a}~ \cite{ZEUS_f_96b}~\cite{ZEUS_f_98}.
}
\end{center}
\end{minipage}
\hfill
\begin{minipage}[t]{0.49\textwidth}
\begin{center}
\psfig{file=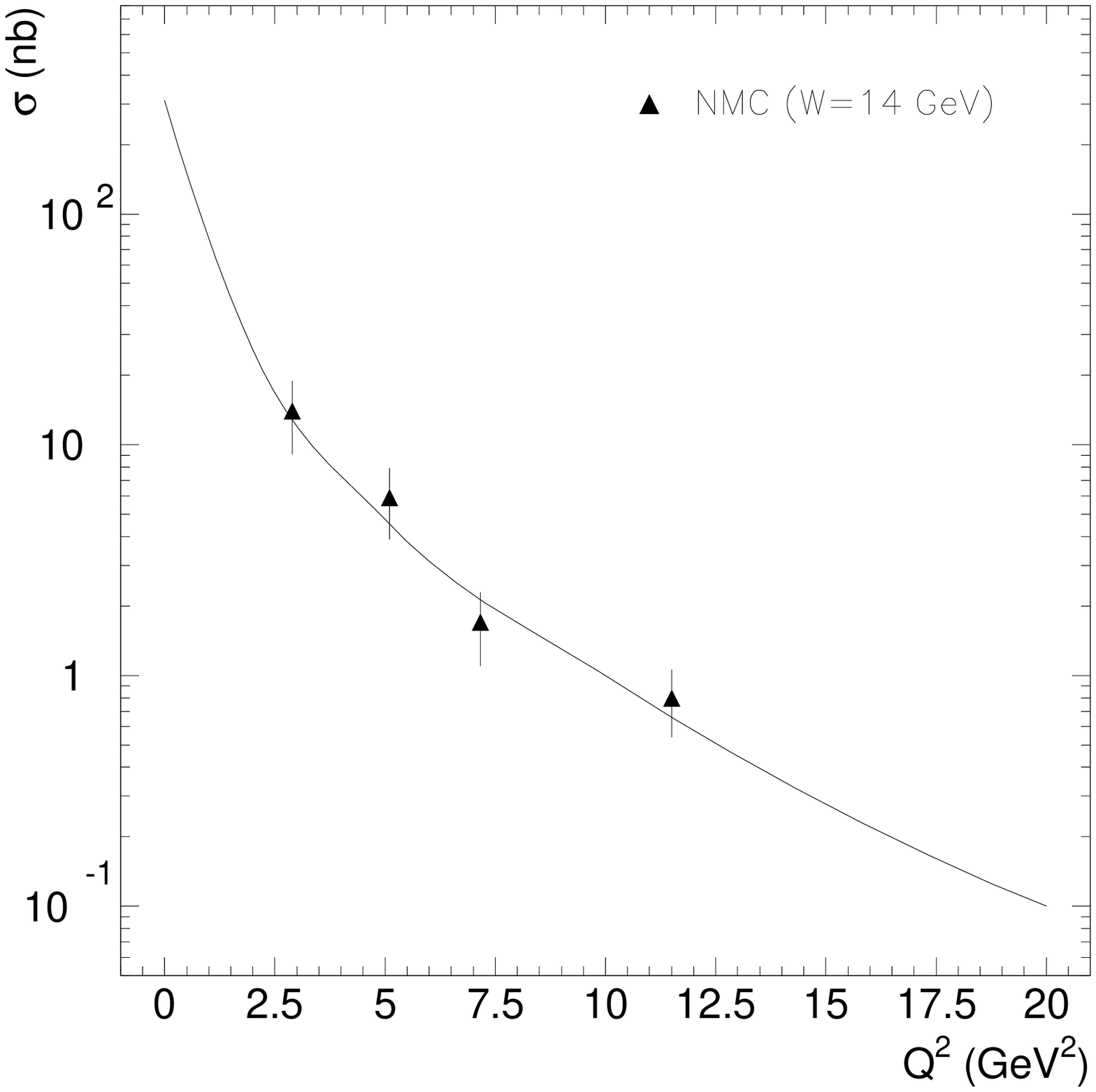, width=\textwidth}
\vspace*{-10mm}
\caption{ 
\label{S1p_b}
$Q^2$ dependence of the \fmes\ cross-section at $W = 14 \, \gev$
in model S1. The data are from NMC  \cite{NMC_r_94}.
}
\end{center}
\end{minipage}
\end{figure}

\begin{figure}[!p]
\vspace*{-5mm}
\begin{minipage}[t]{0.49\textwidth}
\begin{center}
\psfig{file=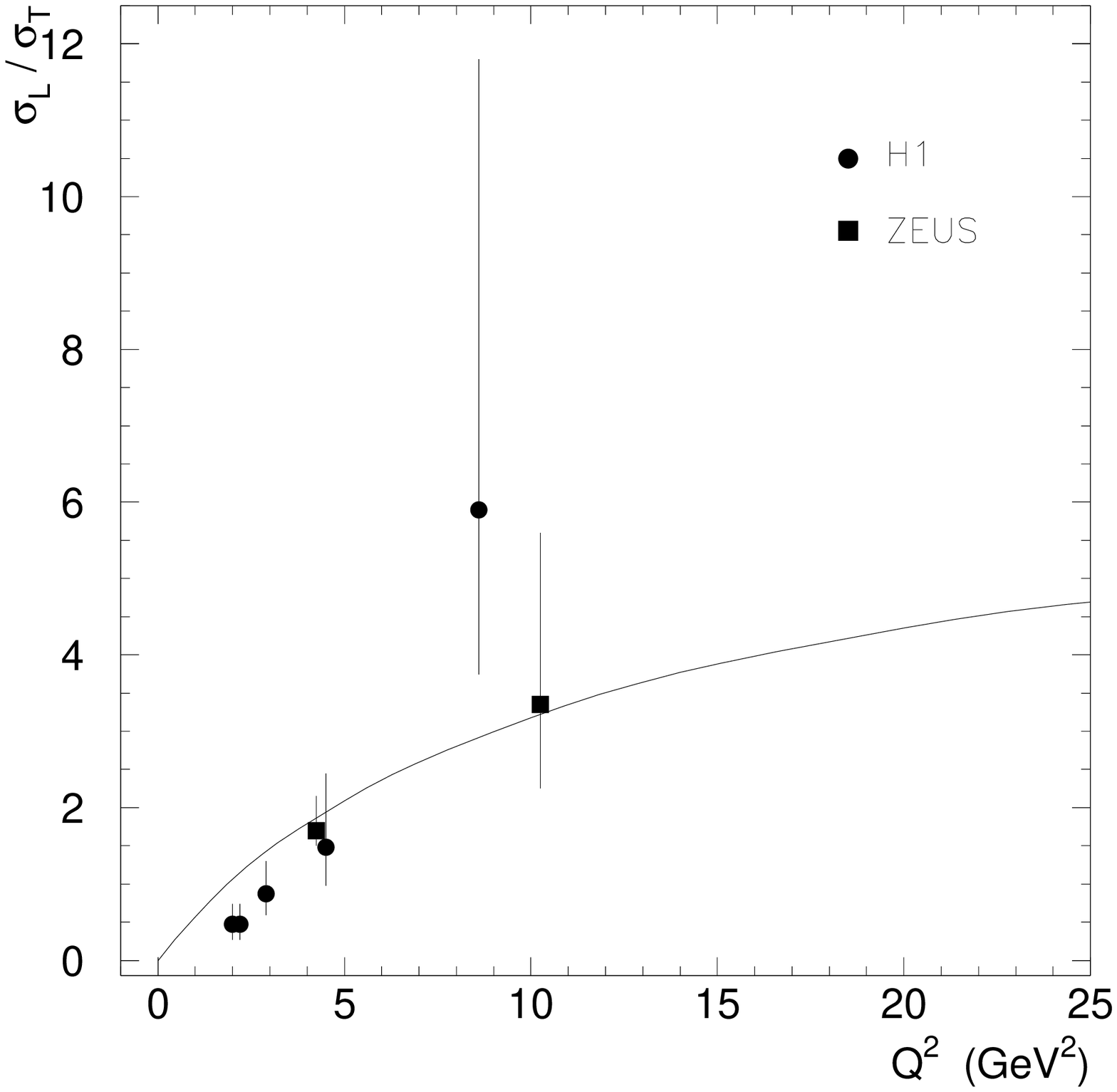, width=\textwidth}
\vspace*{-10mm}
\caption{ 
\label{S1p_c}
$Q^2$ dependence of the \fmes\  longtitudinal to transverse 
cross-section ratio  
at $W = 90 \, \gev$ in model S1. The data are from: H1 \cite{H1_f_00};
and ZEUS  \cite{ZEUS_f_98}.
}
\end{center}
\end{minipage}
\hfill
\begin{minipage}[t]{0.49\textwidth}
\begin{center}
%
%
%
\end{center}
\end{minipage}
\end{figure}

\begin{figure}[!p]
\vspace*{-5mm}
\begin{minipage}[t]{0.49\textwidth}
\begin{center}
\psfig{file=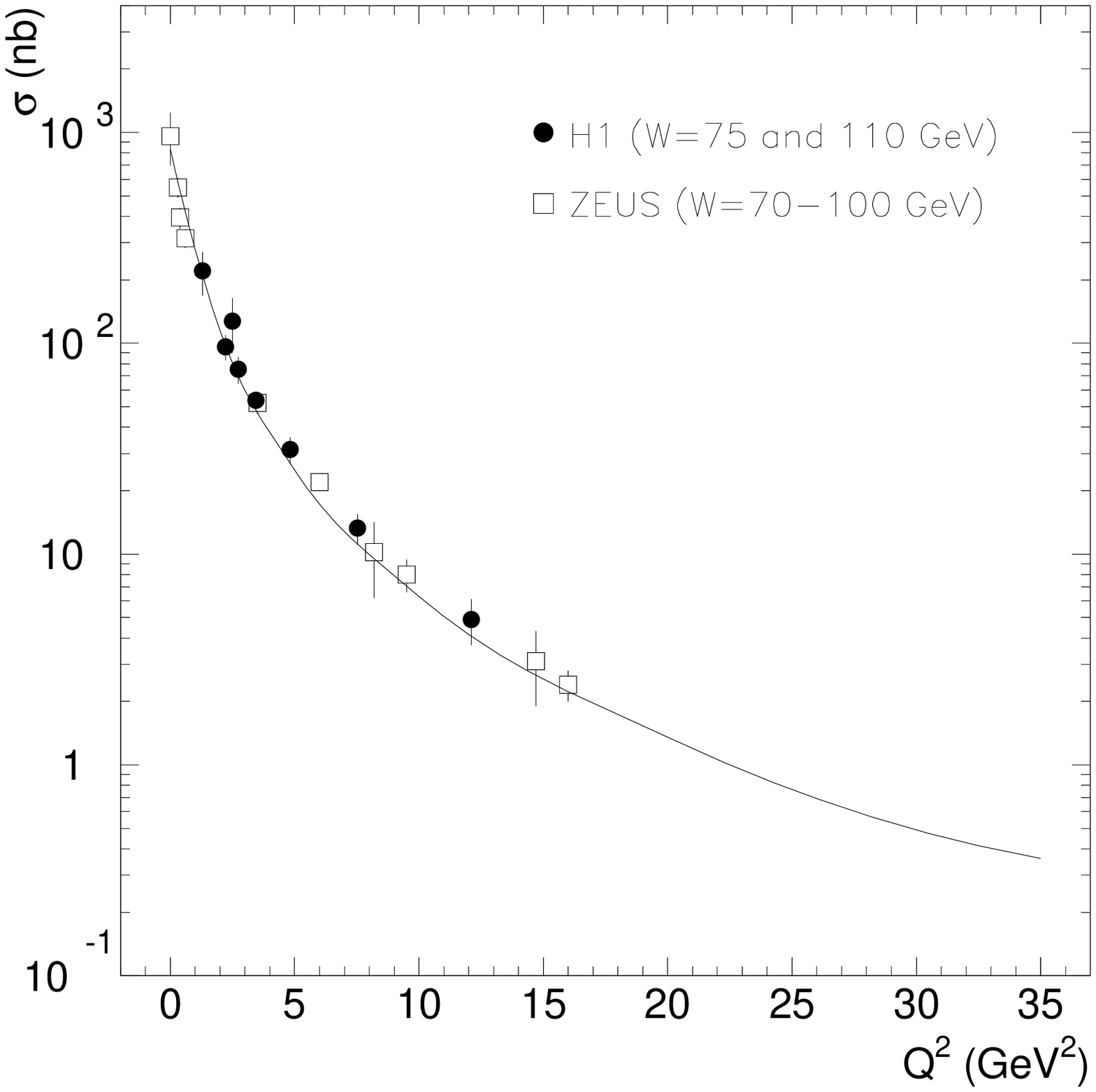, width=\textwidth}
\vspace*{-10mm}
\caption{ 
\label{S2p_a}
$Q^2$ dependence of the \fmes\ cross-section at $W = 90 \, \gev$
in model S2.  The data are from: H1  \cite{H1_f_97}~\cite{H1_f_00};
and ZEUS  \cite{ZEUS_f_96a}~ \cite{ZEUS_f_96b}~\cite{ZEUS_f_98}.
}
\end{center}
\end{minipage}
\hfill
\begin{minipage}[t]{0.49\textwidth}
\begin{center}
\psfig{file=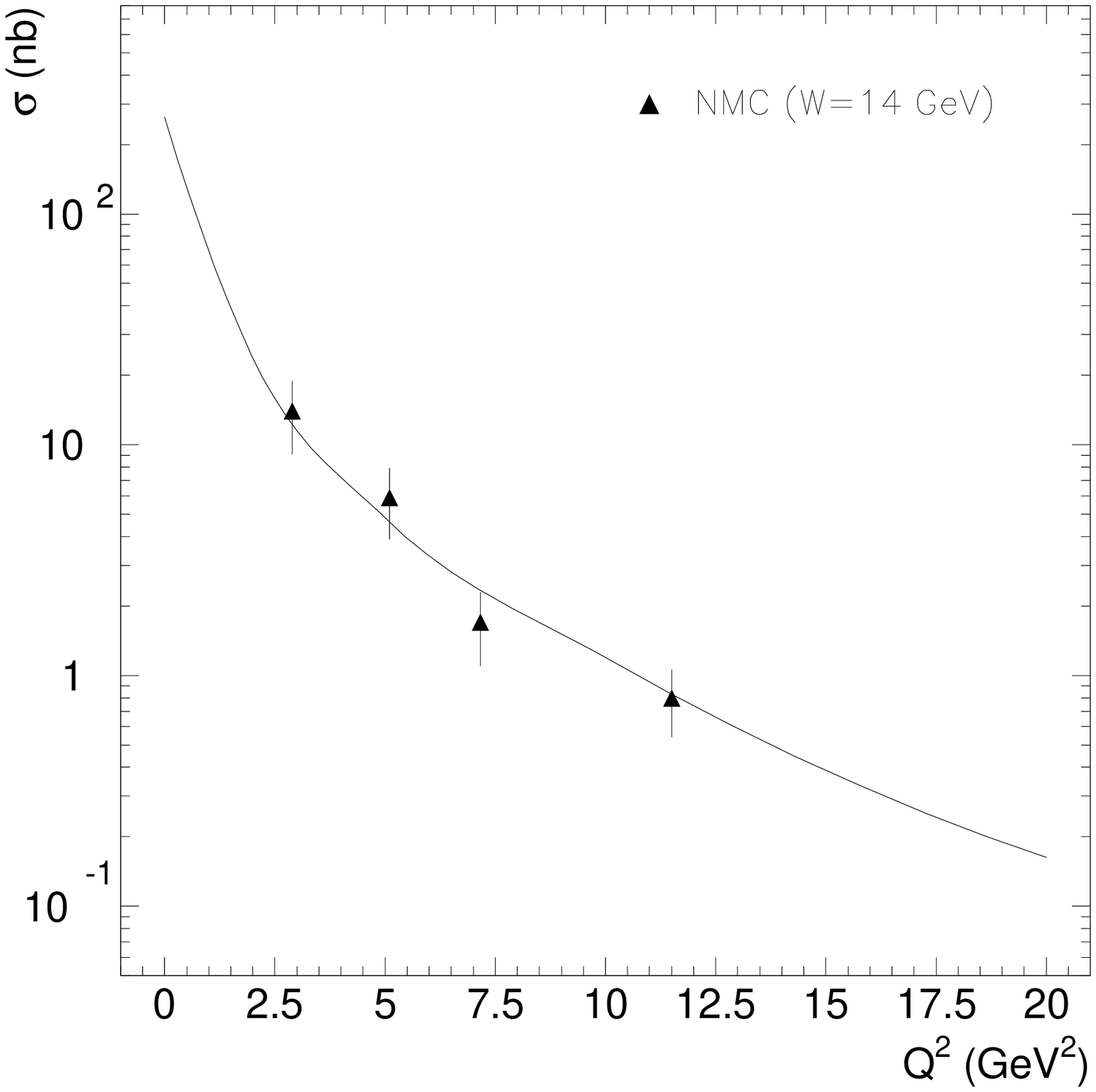, width=\textwidth}
\vspace*{-10mm}
\caption{ 
\label{S2p_b}
$Q^2$ dependence of the \fmes\ cross-section at $W = 14 \, \gev$
in model S2. The data are from NMC \cite{NMC_r_94}.
}
\end{center}
\end{minipage}
\end{figure}

\begin{figure}[!p]
\vspace*{-5mm}
\begin{minipage}[t]{0.49\textwidth}
\begin{center}
\psfig{file=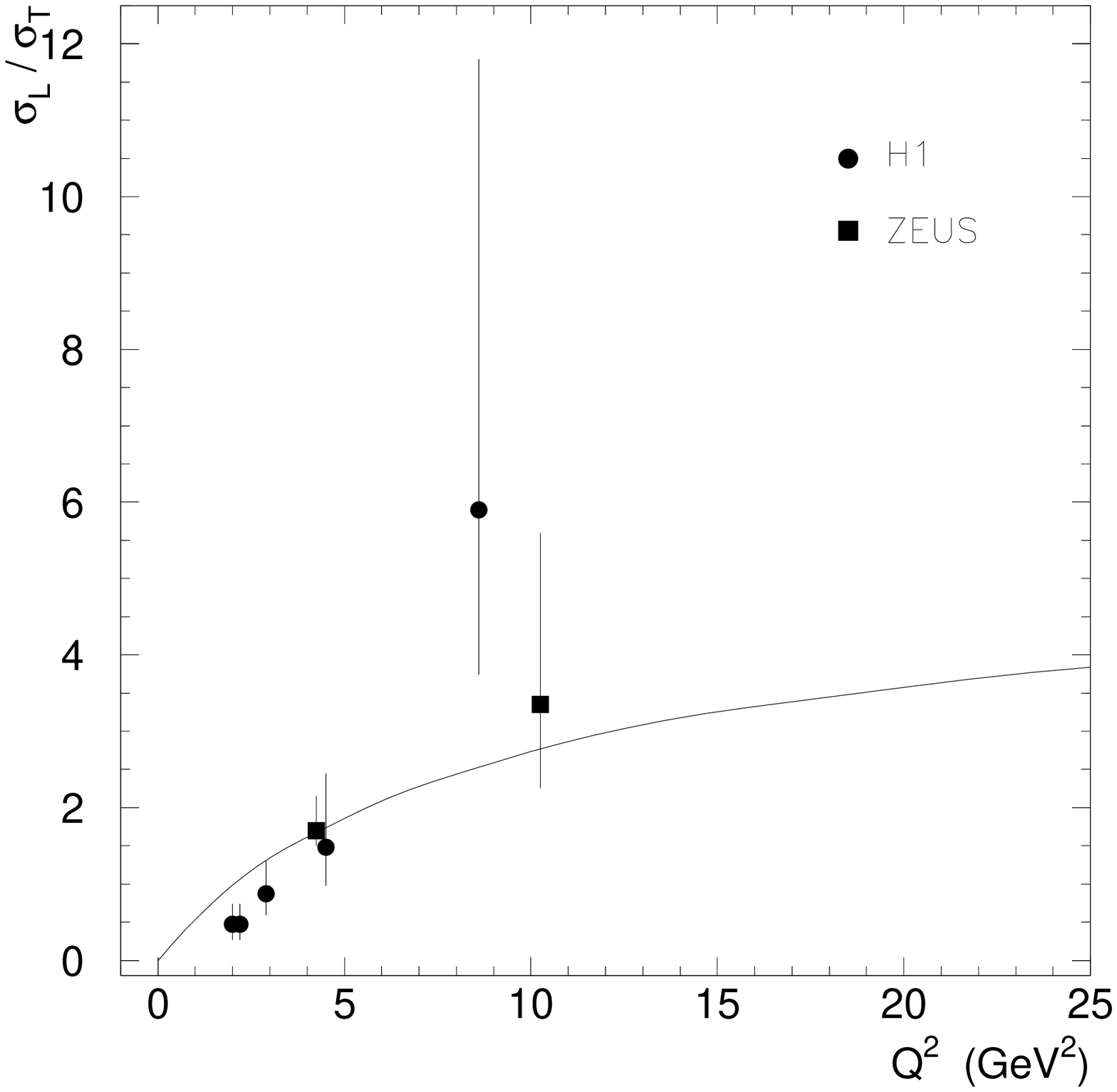, width=\textwidth}
\vspace*{-10mm}
\caption{ 
\label{S2p_c}
$Q^2$ dependence of the \fmes\  longtitudinal to transverse 
cross-section ratio  
at $W = 90 \, \gev$ in model S2.  The data are from: H1 \cite{H1_f_00};
and ZEUS  \cite{ZEUS_f_98}.
}
\end{center}
\end{minipage}
\hfill
\begin{minipage}[t]{0.49\textwidth}
\begin{center}
%
%
%
\end{center}
\end{minipage}
\end{figure}

The results for the $\phi$ with $p_F = 0.6$ GeV are shown in figures 
\ref{S1p_a},
\ref{S1p_b} and \ref{S1p_c} for S1 and in figures \ref{S2p_a}, \ref{S2p_b} 
and \ref{S2p_c} for S2. The results in both cases are 
very good. It is particularly satisfying that the description of the 
low-energy data is good, as in this case there is no reggeon contribution.
Once again the results obtained using the wave functions derived from the
specific potentials are very close to those shown, and provide an equally
good description. 

\begin{figure}[!p]
\vspace*{-5mm}
\begin{minipage}[t]{0.49\textwidth}
\begin{center}
\psfig{file=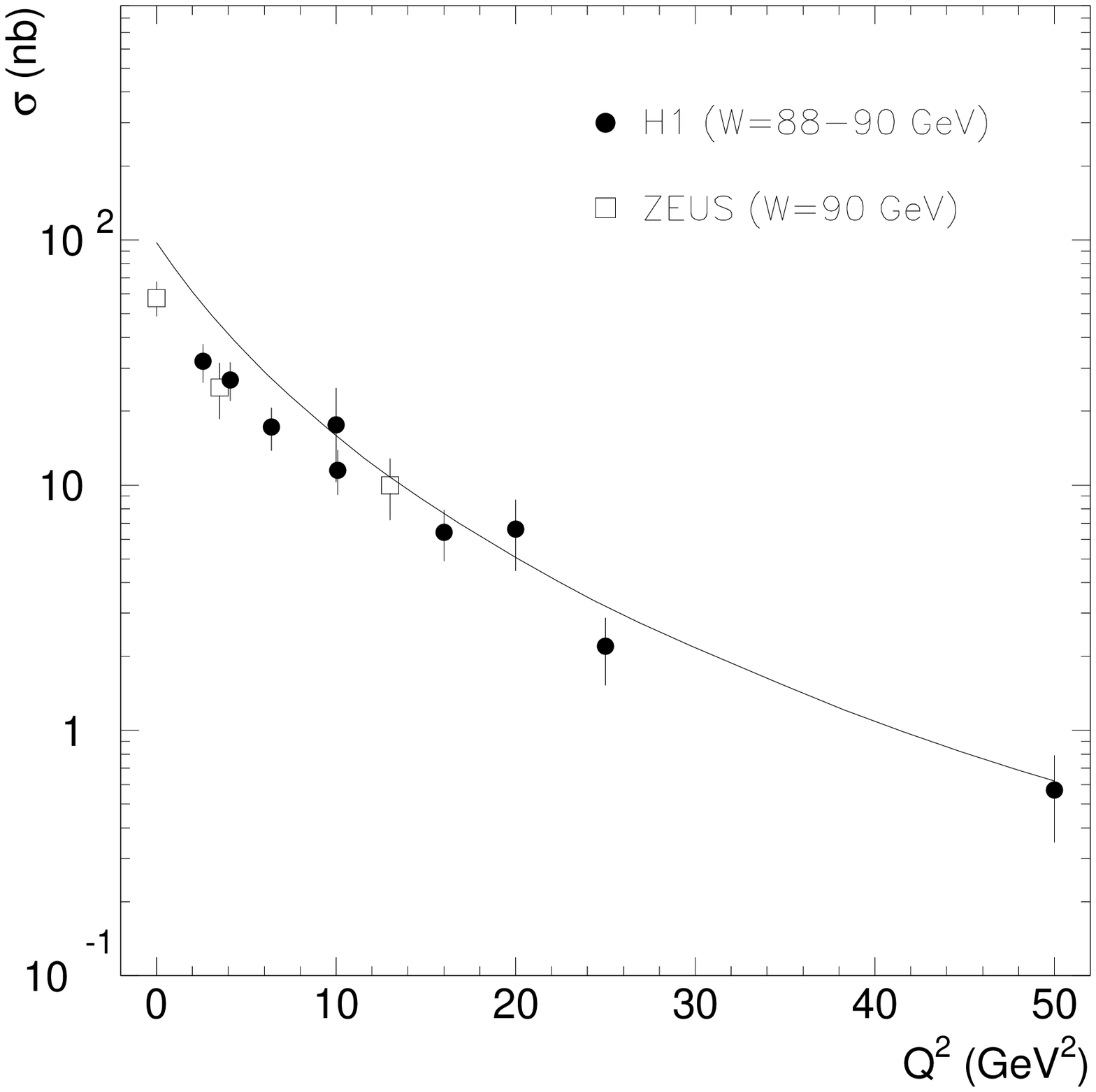, width=\textwidth}
\vspace*{-10mm}
\caption{ 
\label{S1j_a}
$Q^2$ dependence of the \jpsimes\ cross-section at $W = 90 \, \gev$
in model S1. The data are from  H1 \cite{H1_j_96}~\cite{H1_j_99};
and  ZEUS  \cite{ZEUS_j_97}~\cite{ZEUS_r_99}.
}
\end{center}
\end{minipage}
\hfill
\begin{minipage}[t]{0.49\textwidth}
\begin{center}
\psfig{file=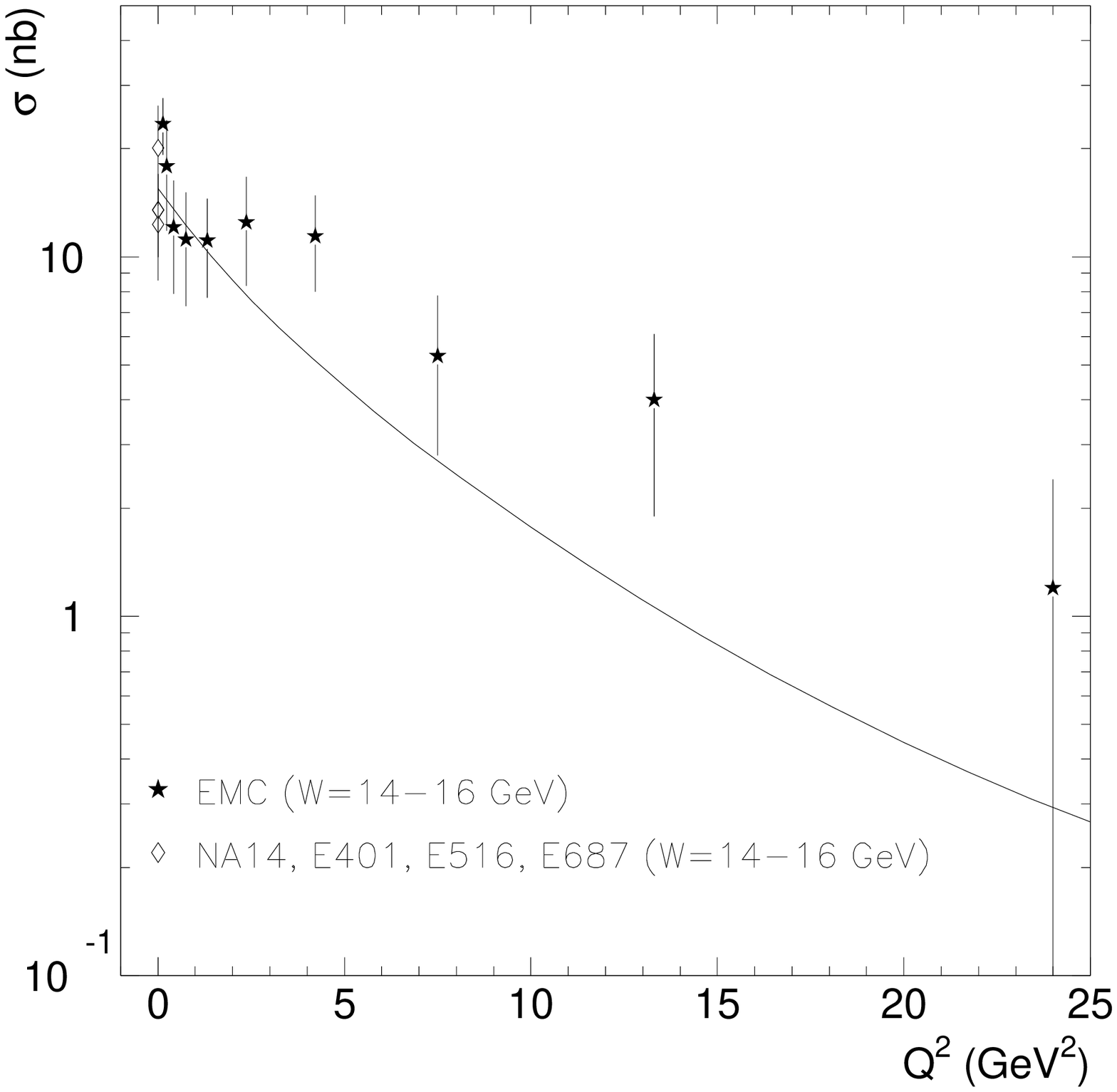, width=\textwidth}
\vspace*{-10mm}
\caption{ 
\label{S1j_b}
$Q^2$ dependence of the \jpsimes\ cross-section at $W = 14 \, \gev$
in model S1. The data are from  EMC  \cite{EMC_j_83};
 E401  \cite{E401_j_82};  E516  \cite{E516_j_84};
 NA14  \cite{NA14_j_87}; and E687  \cite{E687_j_93}.
}
\end{center}
\end{minipage}
\end{figure}

\begin{figure}[!p]
\vspace*{-5mm}
\begin{minipage}[t]{0.49\textwidth}
\begin{center}
\psfig{file=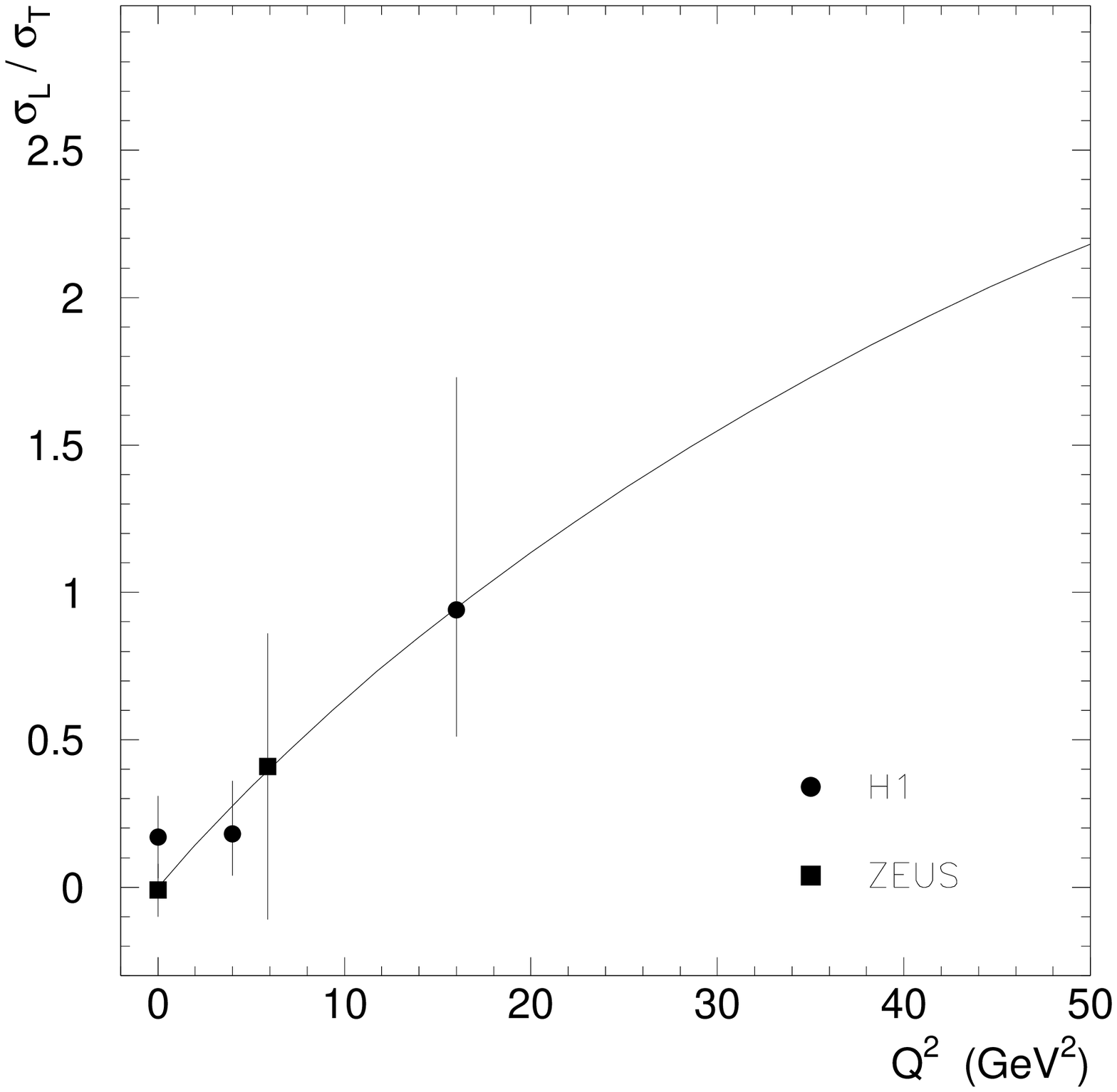, width=\textwidth}
\vspace*{-10mm}
\caption{ 
\label{S1j_c}
$Q^2$ dependence of the \jpsimes\  longtitudinal to transverse 
cross-section ratio  at $W = 90 \, \gev$ in model S1.
The data are from:  H1  \cite{H1_j_96}~\cite{H1_j_99};
and ZEUS  \cite{ZEUS_j_97}~\cite{ZEUS_r_99}.
}
\end{center}
\end{minipage}
\hfill
\begin{minipage}[t]{0.49\textwidth}
\begin{center}
%
%
%
\end{center}
\end{minipage}
\end{figure}

\begin{figure}[!p]
\vspace*{-5mm}
\begin{minipage}[t]{0.49\textwidth}
\begin{center}
\psfig{file=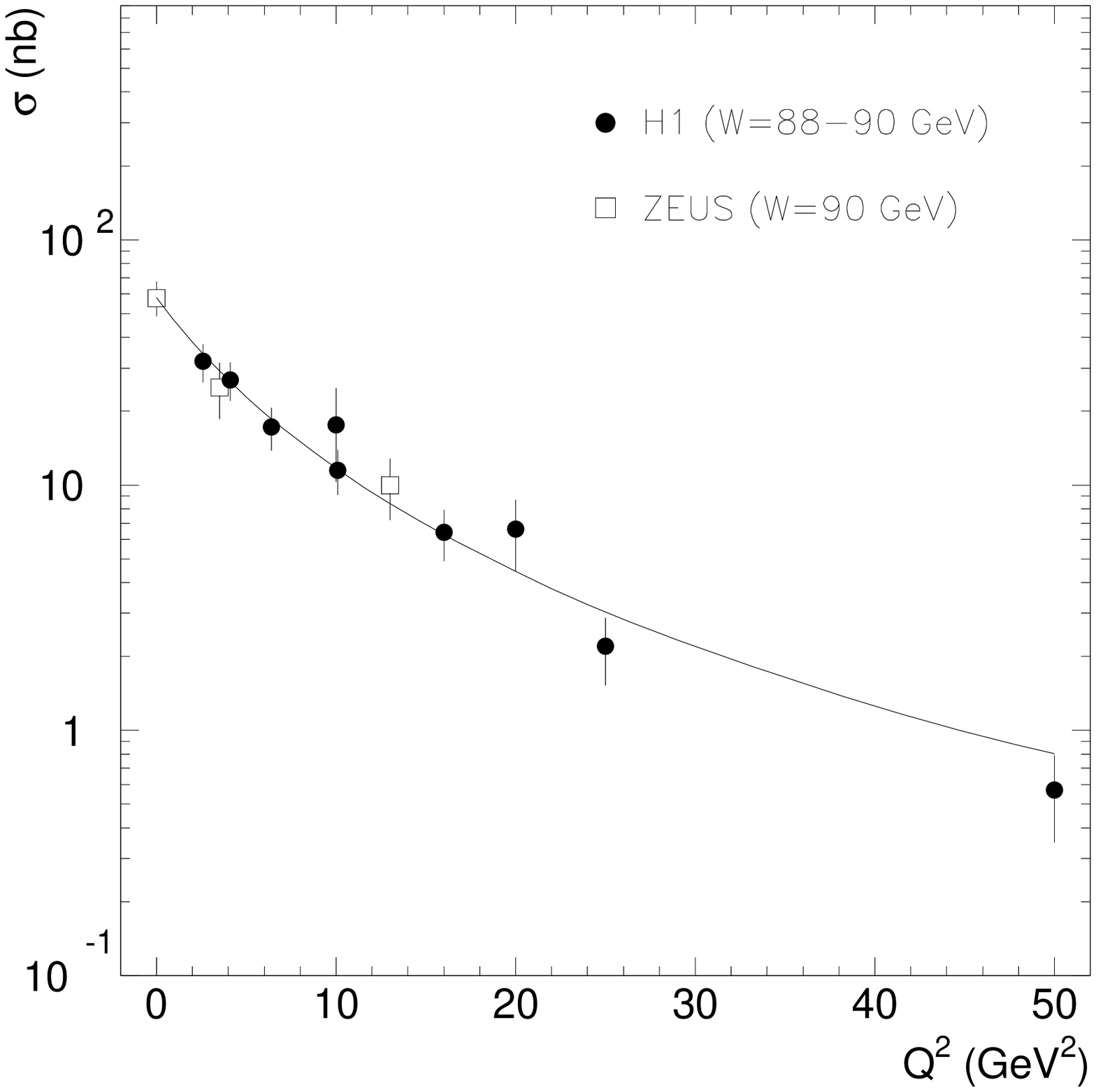, width=\textwidth}
\vspace*{-10mm}
\caption{ 
\label{S2j_a}
$Q^2$ dependence of the \jpsimes\ cross-section at $W = 90 \, \gev$
in model S2.  The data are from  H1 \cite{H1_j_96}~\cite{H1_j_99};
and  ZEUS  \cite{ZEUS_j_97}~\cite{ZEUS_r_99}.
}
\end{center}
\end{minipage}
\hfill
\begin{minipage}[t]{0.49\textwidth}
\begin{center}
\psfig{file=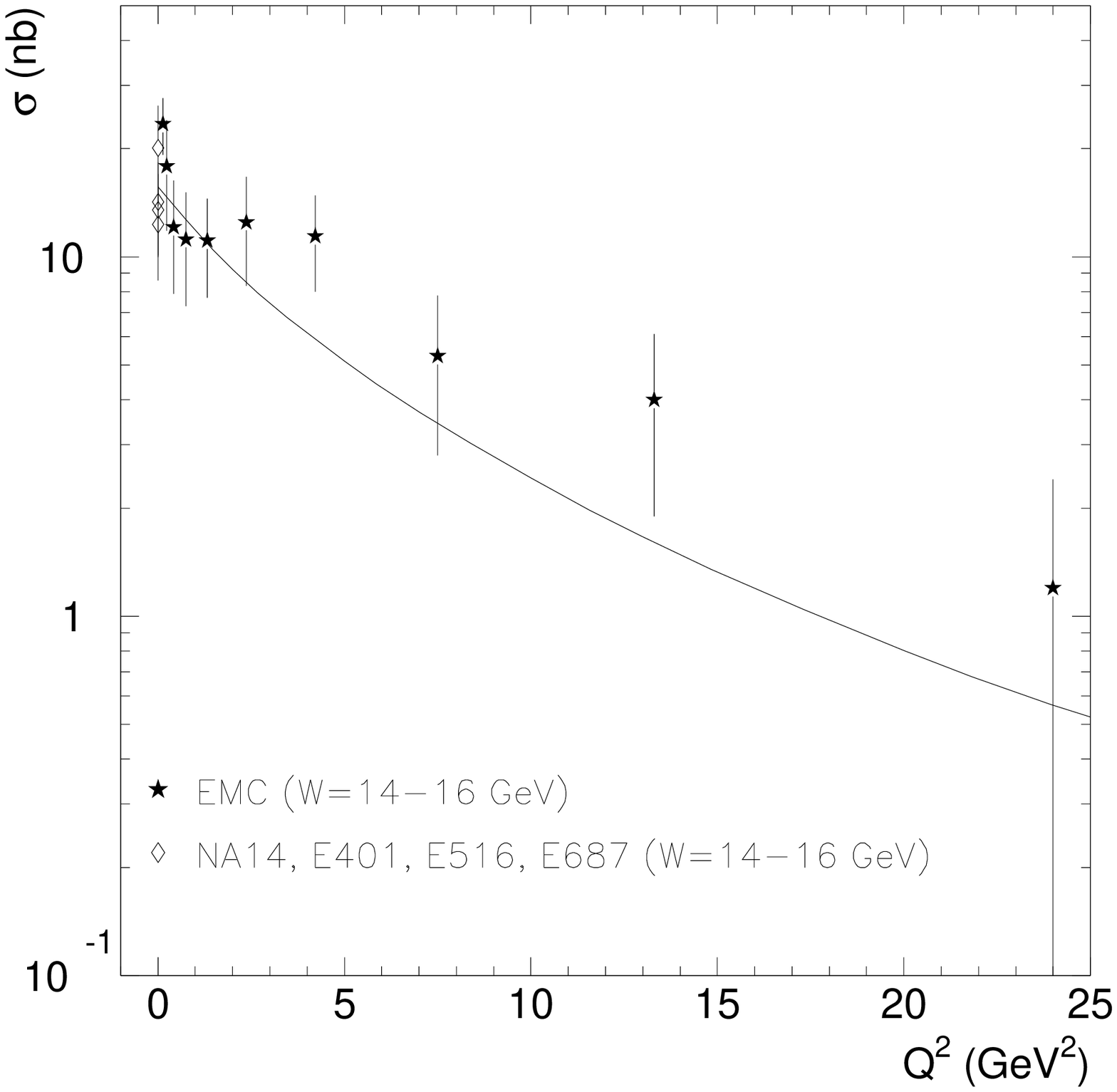, width=\textwidth}
\vspace*{-10mm}
\caption{ 
\label{S2j_b}
$Q^2$ dependence of the \jpsimes\ cross-section at $W = 14 \, \gev$
in model S2. The data are from  The data are from  EMC  \cite{EMC_j_83};
 E401  \cite{E401_j_82};  E516  \cite{E516_j_84};
 NA14  \cite{NA14_j_87}; and E687  \cite{E687_j_93}.
}
\end{center}
\end{minipage}
\end{figure}

\begin{figure}[!p]
\vspace*{-5mm}
\begin{minipage}[t]{0.49\textwidth}
\begin{center}
\psfig{file=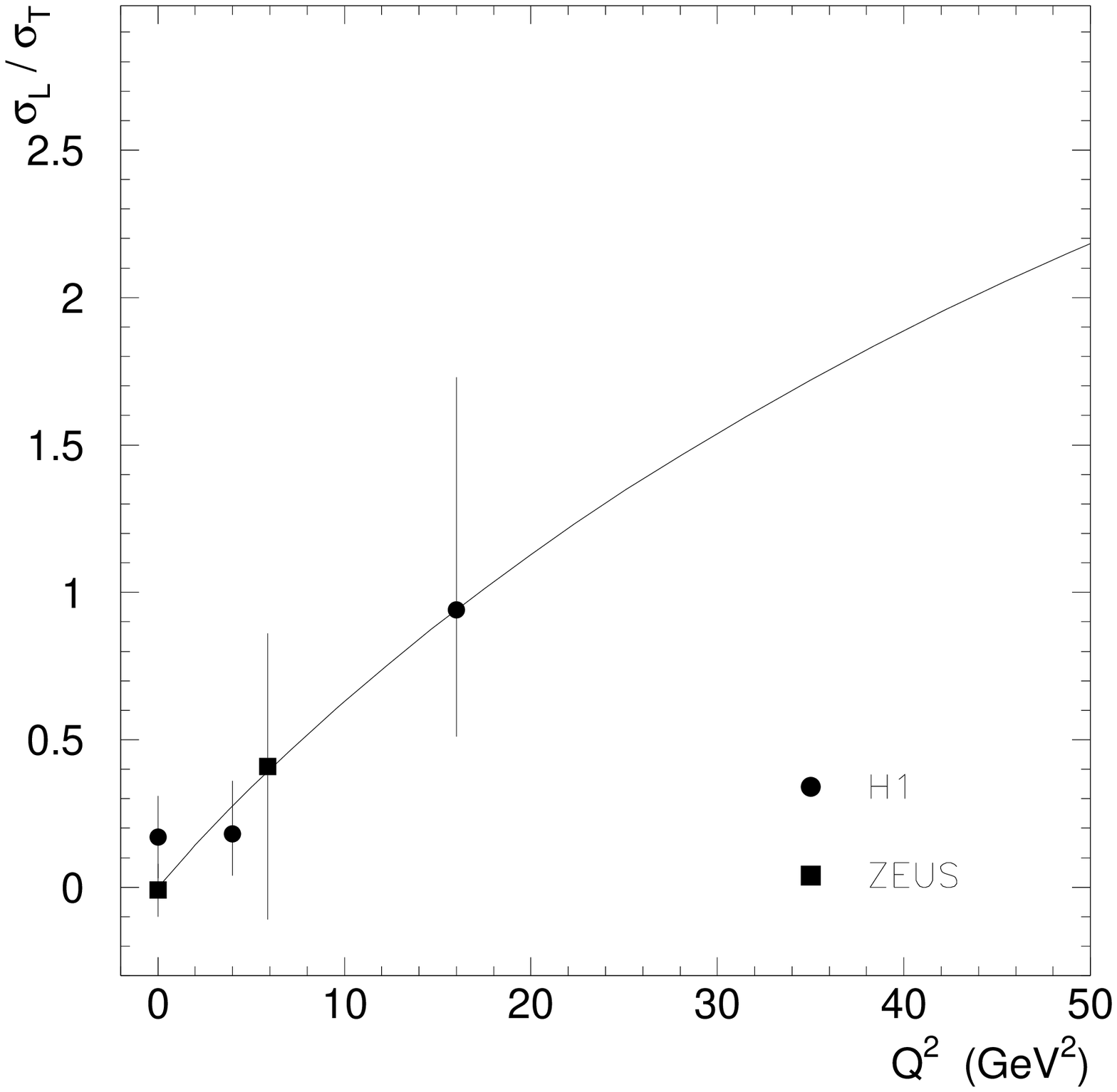, width=\textwidth}
\vspace*{-10mm}
\caption{ 
\label{S2j_c}
$Q^2$ dependence of the \jpsimes\  longtitudinal to transverse 
cross-section ratio  
at $W = 90 \, \gev$ in model S2. The data are from:  H1 
 \cite{H1_j_96}~\cite{H1_j_99};
and ZEUS  \cite{ZEUS_j_97}~\cite{ZEUS_r_99}.
}
\end{center}
\end{minipage}
\hfill
\begin{minipage}[t]{0.49\textwidth}
\begin{center}
%
%
%
\end{center}
\end{minipage}
\end{figure}

The $\jpsi$ results are given in figures \ref{S1j_a} through \ref{S2j_c}
with $p_F =1.2$, and are again very satisfactory. We have included low-energy 
data, although they are not very precise and have some contamination from 
nucleon breakup. 
Because of the mass of the charm quark the data on $\sigma_L/\sigma_T$ ratio 
do not provide a strong constraint, and even at $Q^2 = 50$ GeV$^2$ the ratio
is still far from its asymptotic value. The S2 results are again to be 
slightly preferred overall. In the case of the $\jpsi$ the results obtained 
using the wave functions derived from the specific potentials do not provide
a good description of the data, the cross section being about a factor of
three too large. There are also significant differences in $\sigma_L/\sigma_T$,
the ratio rising almost linearly with increasing $Q^2$, and by $Q^2 = 50$
GeV$^2$ is a factor of two larger than the results shown in figures \ref{S1j_c}
and \ref{S2j_c}. However the present data cannot sensibly distinguish 
between this result and the one shown. That there is a significant 
difference
in the predictions for the $\jpsi$ is not surprising given the very different
wave function used in the fit compared to those obtained from specific
potentials: recall figure \ref{psiwf}. 

\begin{figure}[!p]
\vspace*{-5mm}
\begin{minipage}[t]{0.49\textwidth}
\begin{center}
\psfig{file=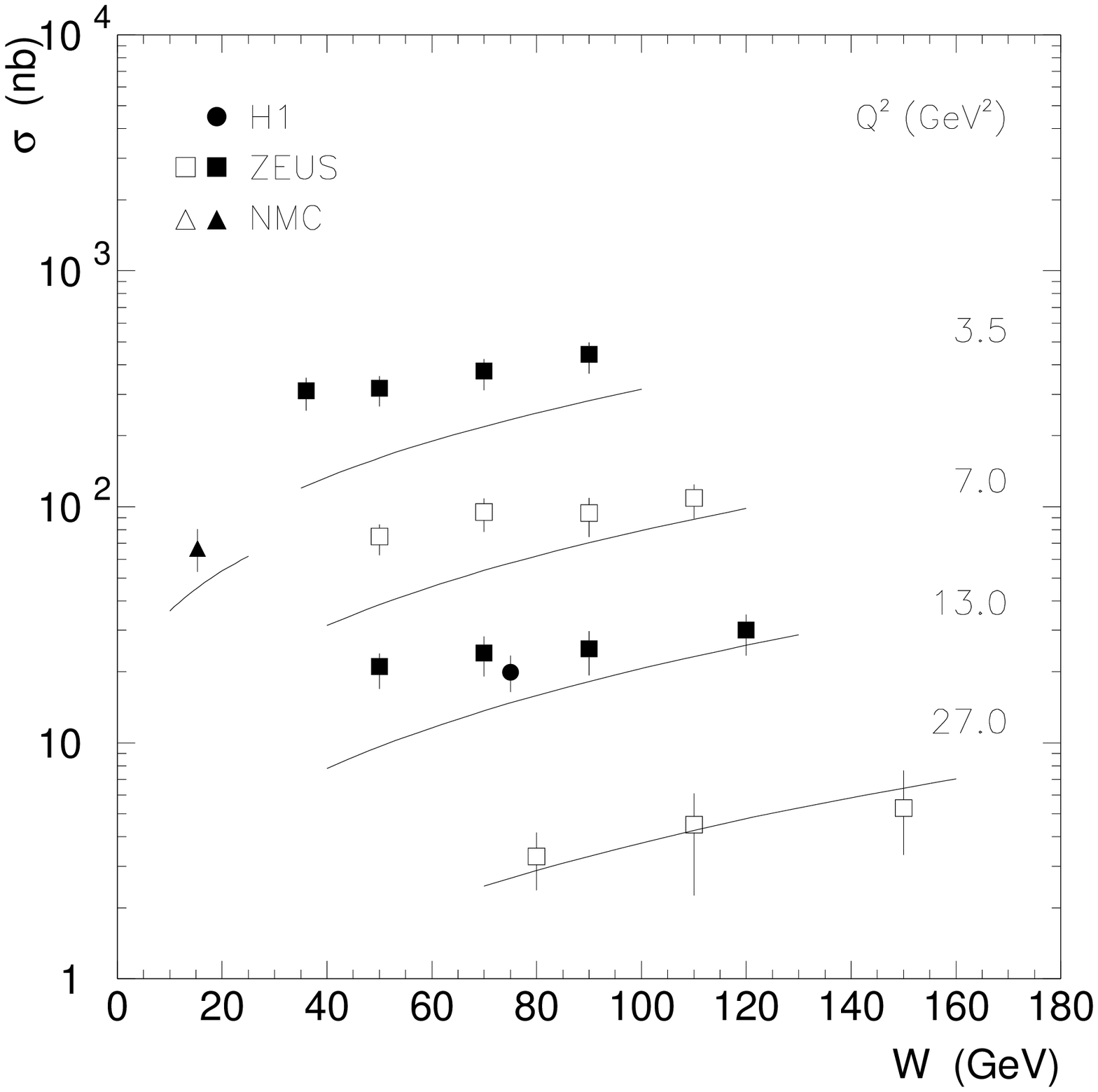, width=\textwidth}
\vspace*{-10mm}
\caption{ 
\label{S2r_e1}
$W$ dependence of the \rmes\ cross-section for various $Q^2$
in model S2. The data are from:  NMC  \cite{NMC_r_94};
 H1  \cite{H1_r_00}; and
 ZEUS  \cite{ZEUS_r_99}. 
}
\end{center}
\end{minipage}
\hfill
\begin{minipage}[t]{0.49\textwidth}
\begin{center}
\psfig{file=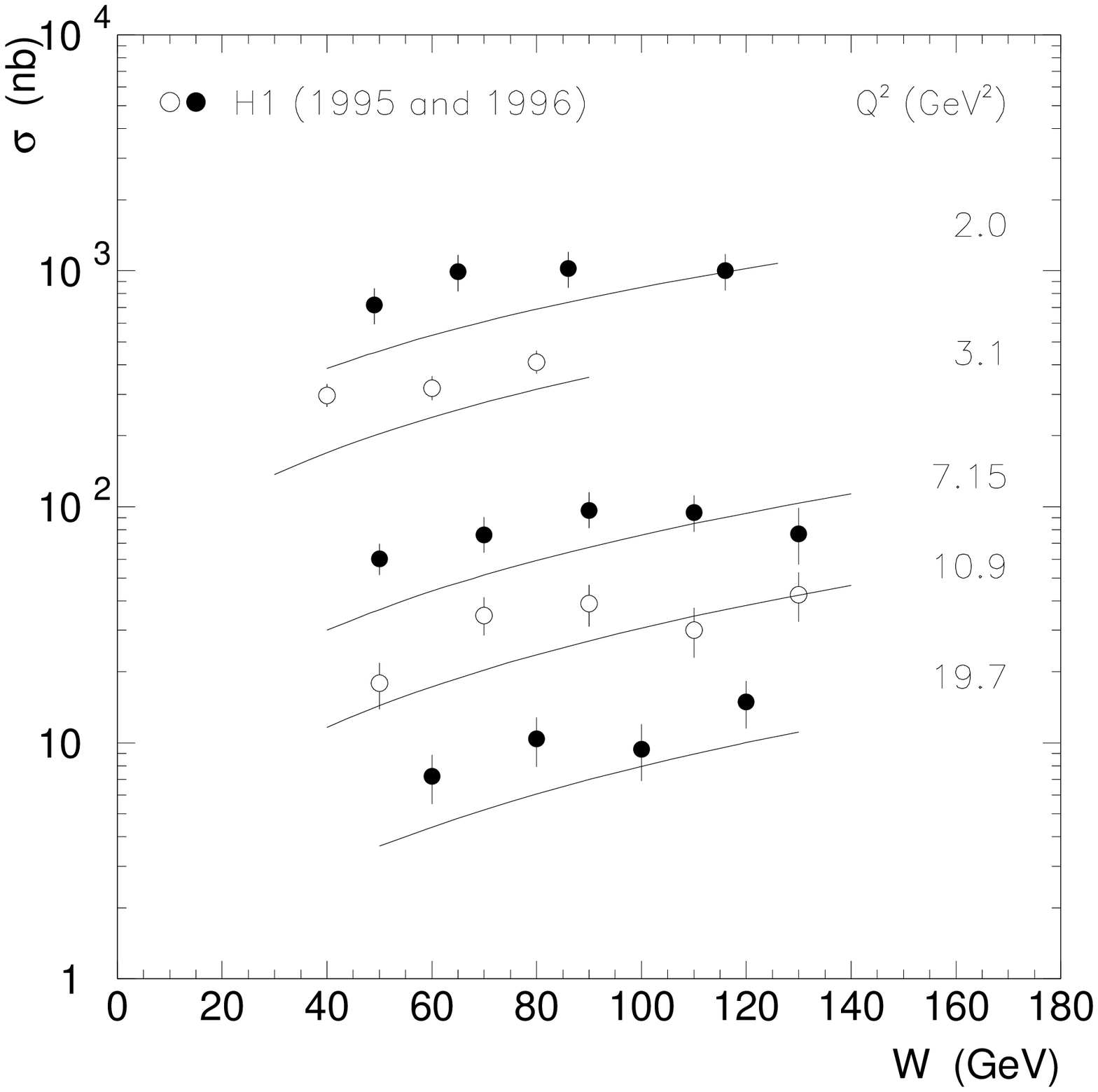, width=\textwidth}
\vspace*{-10mm}
\caption{ 
\label{S2r_e2}
$W$ dependence of the \rmes\ cross-section for various $Q^2$
in model S2. The data are from  H1  \cite{H1_r_00}.
}
\end{center}
\end{minipage}
\end{figure}

\begin{figure}[!p]
\vspace*{-5mm}
\begin{minipage}[t]{0.49\textwidth}
\begin{center}
\psfig{file=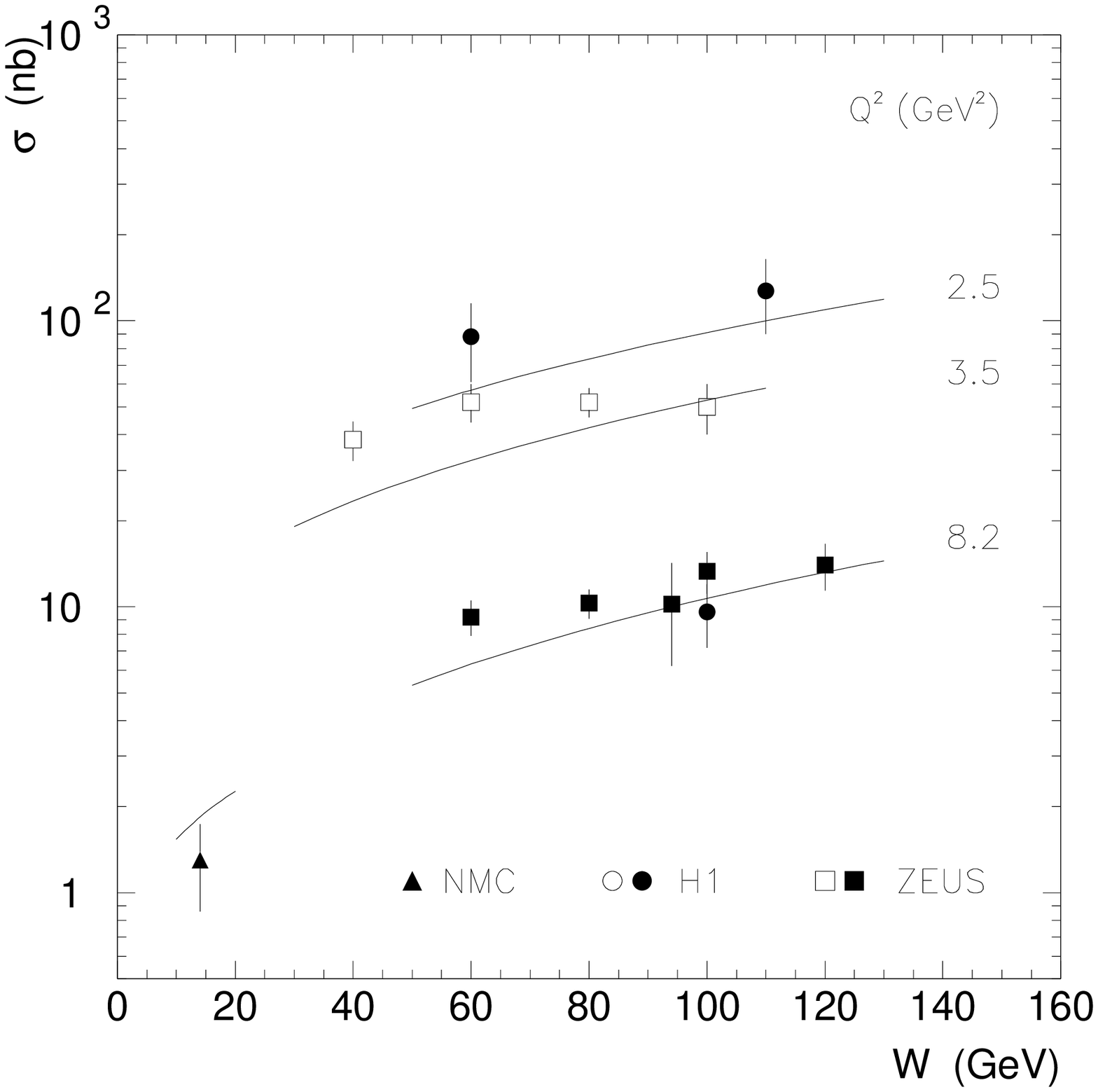, width=\textwidth}
\vspace*{-10mm}
\caption{ 
\label{S2p_e}
$W$ dependence of the \fmes\ cross-section for various $Q^2$
in model S2. The data are from: NMC  \cite{NMC_r_94};
 H1  \cite{H1_f_97};
and ZEUS  \cite{ZEUS_f_96a}~ \cite{ZEUS_f_96b}~\cite{ZEUS_f_98}.
}
\end{center}
\end{minipage}
\hfill
\begin{minipage}[t]{0.49\textwidth}
\begin{center}
\psfig{file=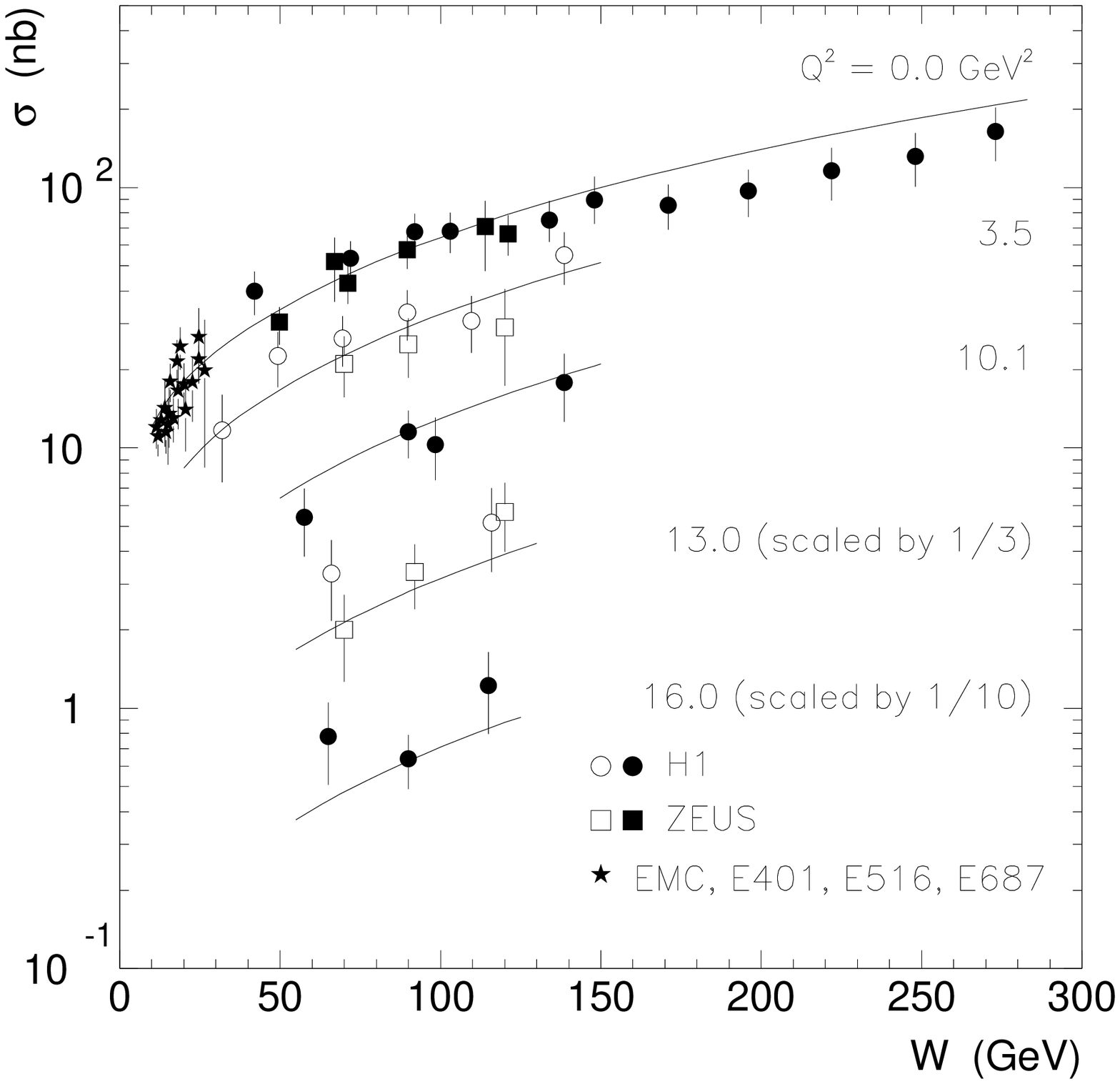, width=\textwidth}
\vspace*{-10mm}
\caption{ 
\label{S2j_e}
$W$ dependence of the $\jpsi$ meson cross-section for various $Q^2$
in model S2. The data are from:  EMC  \cite{EMC_j_83};
 E401  \cite{E401_j_82};  E516  \cite{E516_j_84};
 NA14  \cite{NA14_j_87}; E687  \cite{E687_j_93};
 H1  \cite{H1_j_96}~\cite{H1_j_00}; and
 ZEUS  \cite{ZEUS_j_95}~\cite{ZEUS_j_97}.
}
\end{center}
\end{minipage}
\end{figure}

There is very little difference in the energy dependence predicted by
models S1 and S2, so we shall show only the latter.
The energy dependence at fixed $Q^2$ is shown in figures \ref{S2r_e1}
and \ref{S2r_e2} for the $\rho$, in figure \ref{S2p_e} for the $\phi$ and in
figure \ref{S2j_e} for the $J/\psi$, in each case for the S2 model. The
break in the curves is because of the different value of $\epsilon$ used at 
the lower energies. The model succeeds well in reproducing the trends of the 
data, and is particularly successsful for the $\phi$ and $J/\psi$, even at
$Q^2 = 0$ in the latter case. The increasing energy dependence with 
increasing $Q^2$ is well represented by the model, reflecting the increasing 
importance of the hard pomeron as $Q^2$ is increased. This is not artifically 
imposed on the model. It occurs naturally via the loop integrals involving 
the meson wave function and the gluon propagators. The model also automatically
takes account of the increasing importance of the hard pomeron with increasing
quark mass. For example, for the \r\ at $Q^2 = 0$ the soft pomeron
contributes $95\%$ of the amplitude at $\sqrt{s} = 15$ GeV and $85\%$ of 
the amplitude at $\sqrt{s} = 75$ GeV. At $Q^2 = 20$ GeV$^2$ these have become 
$25\%$ of the amplitude at $\sqrt{s} = 15$ GeV and $10\%$ of the amplitude
at $\sqrt{s} = 75$ GeV. These proportions are very similar for the \f\,
with a slight increase of the hard component: for example at $Q^2 = 0$ the
soft pomeron contributes $90\%$ of the amplitude at $\sqrt{s} = 15$ GeV and
$70\%$ of the ampitude at $\sqrt{s} = 90$ GeV.  For the $\jpsi$ at $Q^2 = 0$
the soft pomeron contributes $80\%$ of the amplitude at $\sqrt{s} = 15$ GeV
and $35\%$ at $\sqrt{s} = 250$ GeV. These results for the soft pomeron 
proportions at $Q^2 = 0$ are comparable with those obtained in other 
phenomenological approaches \cite{DL00,GN00}. The same general features,
a slow onset of the perturbative region where the hard pomeron dominates,
are compatible with those of \cite{Anisovich,FMS}. In  \cite{Anisovich}
a theoretical analysis is made of $\gamma^* p \rightarrow \rho^0 p$
based on the BFKL formalism, and it is concluded that the perturbative
term does not dominate until energies and virtualities beyond those
currently accessed at HERA. In \cite{FMS} an analysis of $J/\psi$
photoproduction in a dipole model clearly illustrates the mixing of
perturbative and nonperturbative effects in this process.

\begin{figure}[!p]
\vspace*{-5mm}
\begin{minipage}[t]{0.49\textwidth}
\begin{center}
\psfig{file=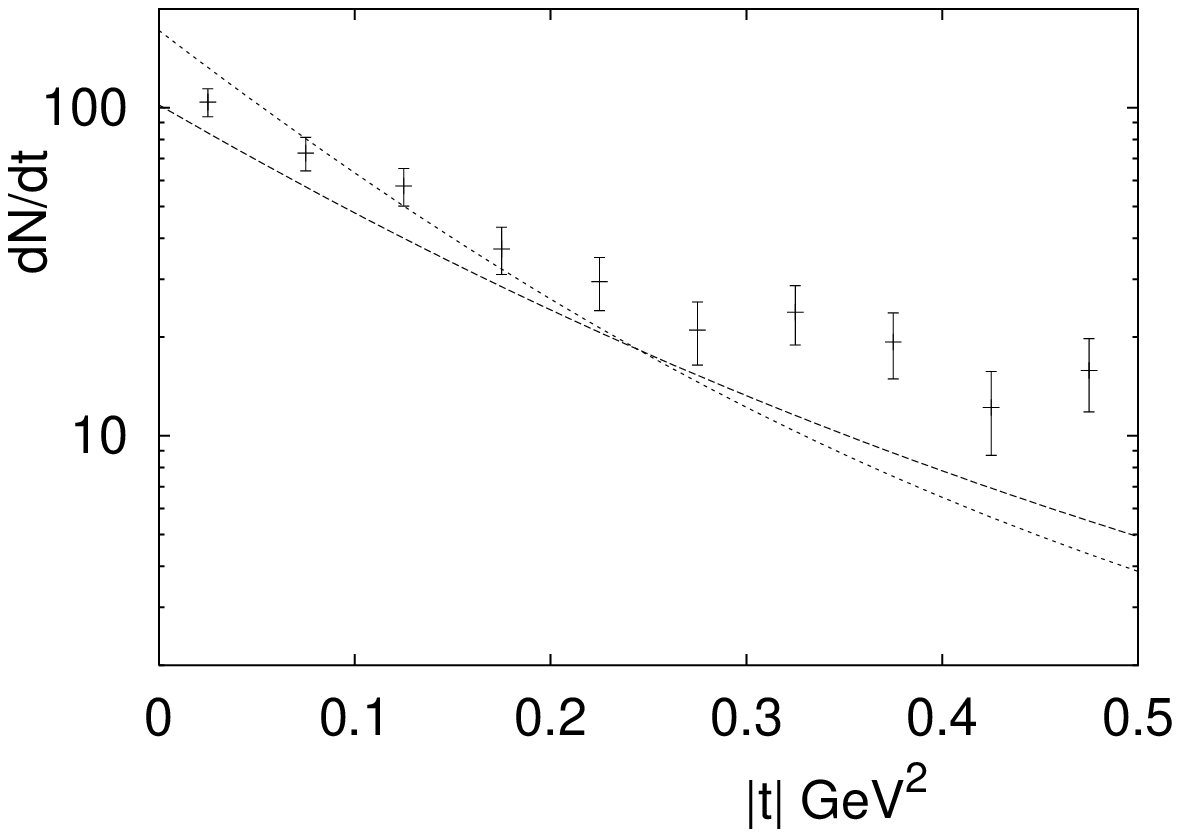, width=\textwidth}
\vspace*{-10mm}
\caption{ 
\label{rho_t1}
$\rho$ meson $t$ dependence, $\langle Q^2 \rangle = 4.8$ GeV$^2$, 
$30 < W < 140$ GeV.
The dashed line is for the S1 model, the dotted line for the S2 model.
The overall normalisation is arbitrary, but the relative normalisation
of the S1 and S2 predictions are as given by the model. 
The data are from H1 \cite{H1_r_00}.
}
\end{center}
\end{minipage}
\hfill
\begin{minipage}[t]{0.49\textwidth}
\begin{center}
\psfig{file=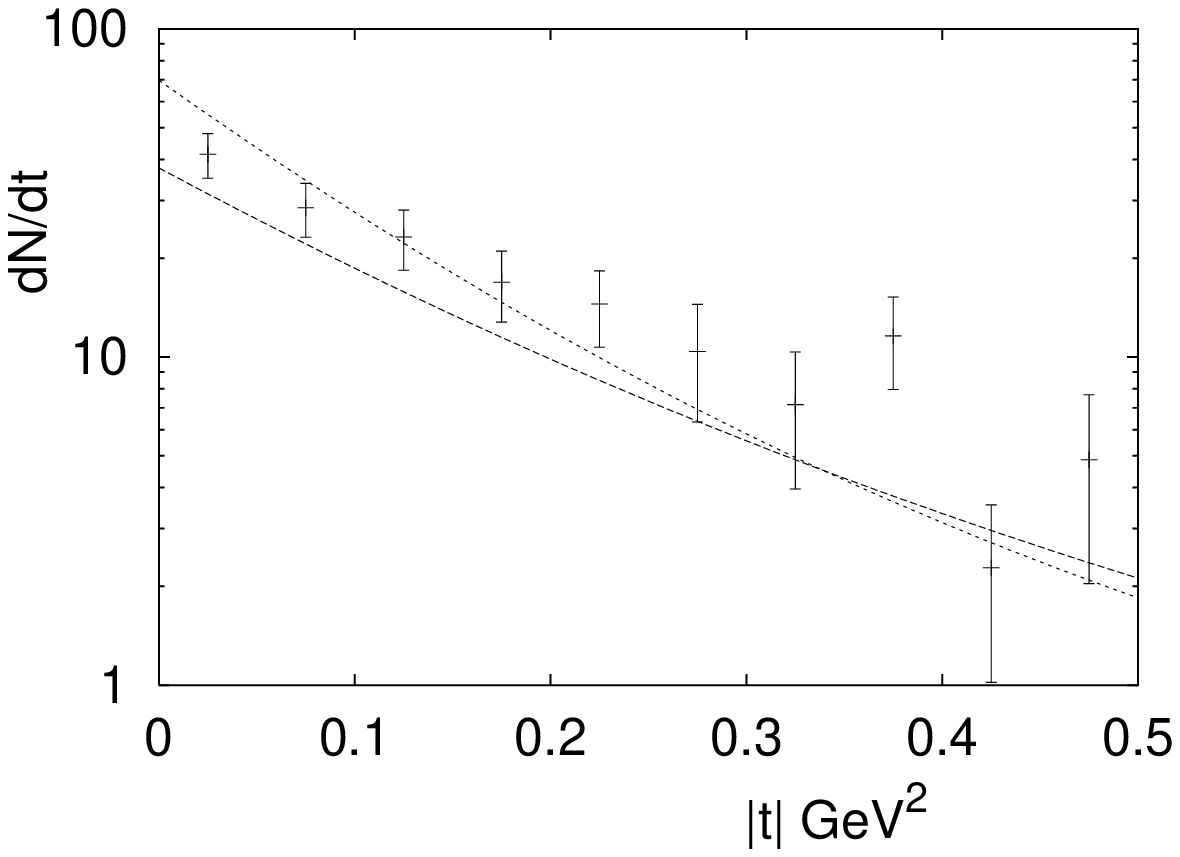, width=\textwidth}
\vspace*{-10mm}
\caption{ 
\label{rho_t2}
$\rho$ meson $t$ dependence, $\langle Q^2 \rangle = 10.9$ GeV$^2$,
$30 < W < 140$ GeV.
The dashed line is for the S1 model, the dotted line for the S2 model.
The overall normalisation is arbitrary, but the relative normalisation
of the S1 and S2 predictions are as given by the model. The data are from
 H1 \cite{H1_r_00}.
}
\end{center}
\end{minipage}
\end{figure}

\begin{figure}[!p]
\vspace*{-5mm}
\begin{minipage}[t]{0.49\textwidth}
\begin{center}
\psfig{file=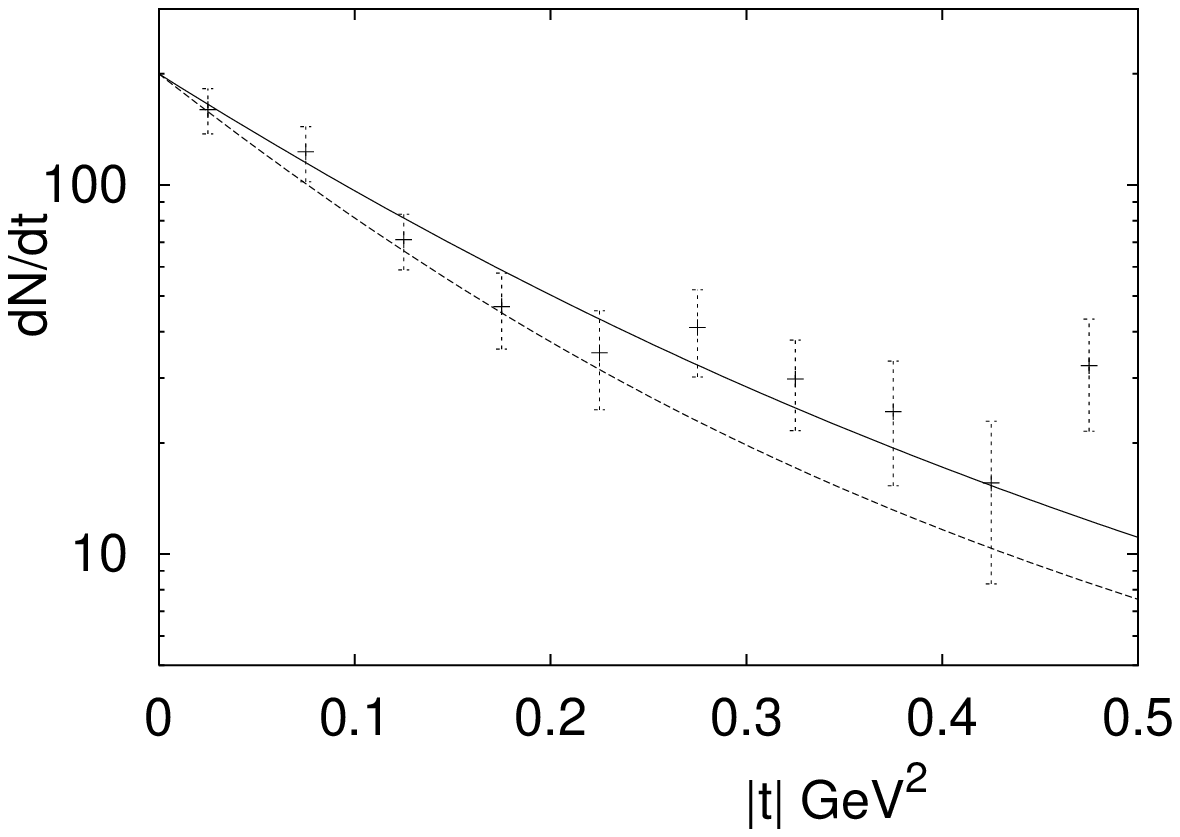, width=\textwidth}
\vspace*{-10mm}
\caption{ 
\label{phi_t}
$\phi$ meson $t$ dependence, $\langle Q^2 \rangle = 4.5$ GeV$^2$,
$40 < W < 130$ GeV.
The dashed line is for the S1 model, the dotted line for the S2 model.
The overall normalisation is arbitrary, but the relative normalisation
of the S1 and S2 predictions are as given by the model.
The data are from  H1  \cite{H1_f_00}.
}
\end{center}
\end{minipage}
\hfill
\begin{minipage}[t]{0.49\textwidth}
\begin{center}
\psfig{file=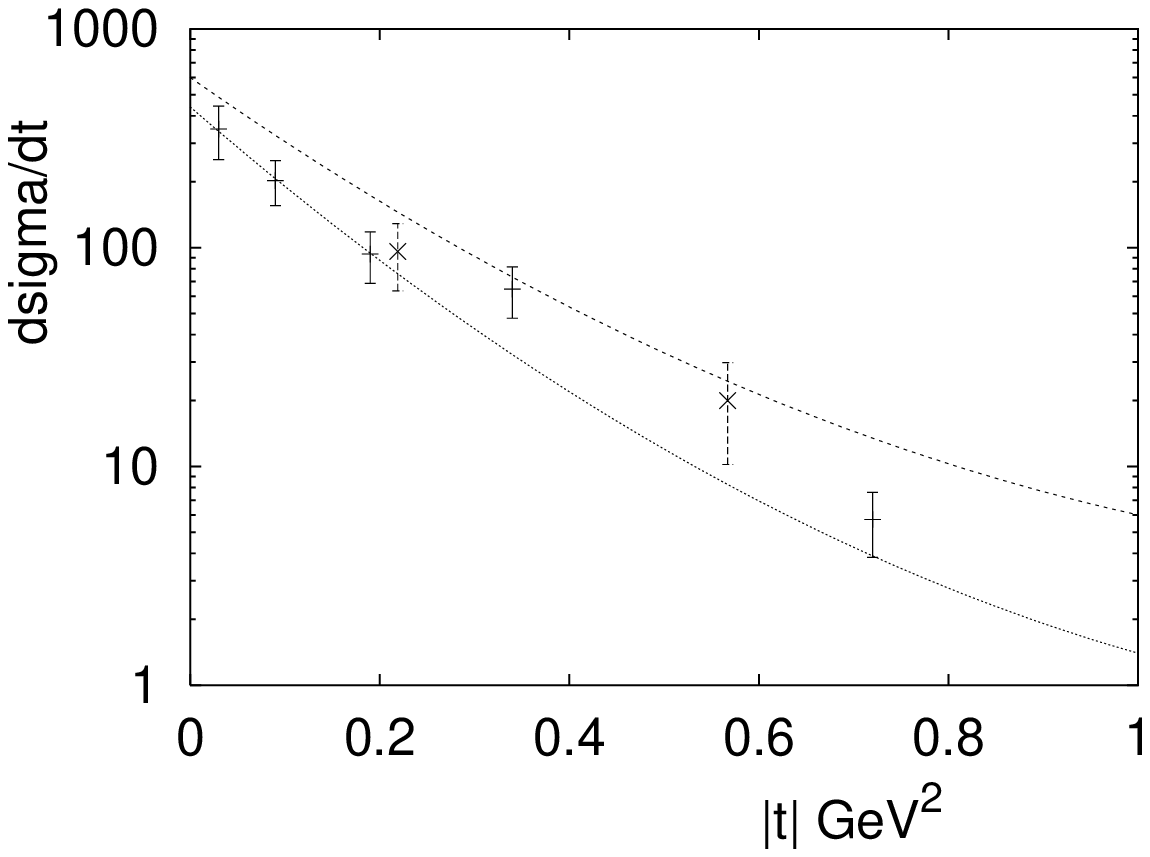, width=\textwidth}
\vspace*{-10mm}
\caption{ 
\label{psi_t}
$J/\psi$ meson $t$ dependence, $Q^2 = 0$, $91 < W < 110$ GeV.
The dashed line is for the S1 model, the dotted line for the S2 model.
The data are from:  H1  \cite{H1_j_00};
and ZEUS  \cite{ZEUS_j_00}. 
}
\end{center}
\end{minipage}
\end{figure}

Finally, examples of the predicted $t$-dependence are shown in 
figures \ref{rho_t1} to 
\ref{psi_t}. The data for the \r\ and \f\ are unnormalised, so the
theoretical curves have been renormalized to the data. The relative 
normalisation of S1 and S2
is that given by the models. For the $\jpsi$ the data and the predictions
are normalised. Clearly the description is again very satisfactory,
with the possible exception of the \r\, where the models predict a somewhat
faster decrease with $t$ than is observed.


\section{Conclusions}
We have presented a model of vector meson electroproduction, and also for
photoproduction in the case 
of the $\jpsi$. The model gives a good overall description of
the data for $\rho$, $\phi$ and $J/\psi$ electroproduction with 
only five adjustable parameters. The dynamical mechanism, with two 
adjustable parameters, is common
to each and the only freedom in going from one vector meson to another is
the parameter $p_F$ which is related to the ``size'' of the meson in its
rest frame. The values of $p_F$ required for each of the $\rho$, $\phi$ 
and $J/\psi$ are 0.5 GeV$^2$, 0.6 GeV$^2$ and 1.2 GeV$^2$ respectively,
in accord with what one would expect. Not surprisingly, electroproduction
of the $\rho$ is primarily non-perturbative at small $Q^2$, and an important 
non-perturbative component is still present at the highest energy and largest 
$Q^2$ for which data exist. In contrast the perturbative contribution
dominates in $J/\psi$ production at high energy, although at small $Q^2$
interference with the non-perturbative contribution remains important.
For $W \ge 50$ GeV, $Q^2 \ge 20$ GeV$^2$ $\jpsi$ electroproduction can be 
considered to be exclusively perturbative. We finally note that this type
of model could be used to give an excellent description of any of the 
$\rho$, $\phi$ and $J/\psi$ if they were to be considered in isolation.

\section*{Acknowledgments}

This work was supported by the Overseas Research Students Award,
a University of Manchester Research Scholarship and by PPARC grant
number PPA/G/0/1998.


\newpage

%
%
%
%
%
%
%
%
%
%
%
%
%
%
%
%
%
%
%
%

%
%
%
%


\end{document}